\documentclass[%
 reprint,
 amsmath,amssymb,
 aps,
]{revtex4-1}
\usepackage[final,dvips]{graphicx}
\usepackage{verbatim}
\usepackage[utf8]{inputenc}
\usepackage{mathtools}
\usepackage{framed}
\usepackage{braket}
\usepackage{amsmath,amssymb,latexsym}
\usepackage{wrapfig}
\usepackage{subfig}
\usepackage{times}
\usepackage{epsfig}
\usepackage{verbatim}
\usepackage{graphicx}
\usepackage{color}
\usepackage{hyperref}

\date{\today}

\newcommand{\br}{\textbf{r}}
\newcommand{\bk}{\textbf{k}}

\newcommand{\bx}{\textbf{x}}
\newcommand{\bxp}{\textbf{x}^{\prime}}
\newcommand{\be}{\begin{equation}}
\newcommand{\ee}{\end{equation}}

\newcommand \E[1] {Eq.~(\ref{#1})}
\newcommand \EE[1] {Equation~(\ref{#1})}
\newcommand \Es[1] {Eqs.~(\ref{#1})}
\newcommand \Ess[2] {Eqs.~(\ref{#1})-(\ref{#2})}
\newcommand \ben {\begin{eqnarray}}
\newcommand \een {\end{eqnarray}}
\newcommand \bea {\begin{eqnarray}}
\newcommand \eea {\end{eqnarray}}

\begin{document}
\title{Thermo-Density Coupling in PFC Type Models for the Study of Rapid Crystallization}
\author{Gabriel Kocher}
\affiliation{Department of Physics, Centre for the Physics of Materials, McGill University, Montreal, QC, Canada}
\author{Nikolas Provatas}
\affiliation{Department of Physics, Centre for the Physics of Materials, McGill University, Montreal, QC, Canada}

\begin{abstract}
We self-consistently derive a formalism that couples a Phase Field Crystal (PFC) density field to thermal transport. It yields a theory for non-uniform transient temperature and density evolution, and includes local latent heat release during atomic rearrangements of the PFC density field.
The  basic formalism is applied to the original PFC model, demonstrating its capacity to capture heat transfer and recalescence in solidification.  With an aim towards consistently incorporating temperature and other thermodynamic variables into PFC modelling, a new classical density field theory for solid/liquid/vapor systems is then derived. It presents a different approach to those  used in the PFC literature while retaining the major advantages that have become the hallmark of PFC modelling; the new model is also based entirely on physical density, temperature and pressure scales. We end the paper by applying the thermal-density coupling formalism to this new multi-phase density functional theory/PFC model.
\end{abstract}

\maketitle 

\section{Introduction}

Solidification is a classic paradigm for understanding pattern formation in materials. The competition between thermodynamic driving forces, mass and heat transfer and interface energy control the scale and morphology of  microstructure in materials \cite{Lan80}. Beyond its fundamental importance to self-assembly and non-equilibrium thermodynamics, solidification is thus also of practical importance as it establishes many important properties of engineering materials \cite{dantzig2009b}. Traditional theories of solidification are typically designed around the assumption that the interface is in local equilibrium, an assumption that is valid in processes operating at low solidification rates. Emerging technologies, such as those using local laser welding to additively assemble components, typically operate at rapid solidification rates \cite{Olakanmi15,Acharya17}. Here, the cooling rates involved are so large that the notion of interface equilibrium is no longer valid. This leads to the production of metastable solidified states due to solute and density trapping, and kinetic-limited morphologies that are very different from the standard dendritic forms in slow-cooling processes. Moreover, thermally-induced stresses can lead to exotic atomic-scale effects such as defect migration, void formation and precipitation of second phases near stressed interfaces \cite{dantzig2009b,Divya16,Martin2017}. 

Modelling the wide scope of phenomena involved in rapid solidification self-consistently is challenging due to the multiple length and time scales involved. Molecular dynamics is typically limited to times scales that preclude processes active on diffusional time scales, and in systems covering several microns. On the other end of the spectrum, traditional phase field (PF) models can capture diffusional time scales over many hundreds of microns. Their use in quantitative modelling of solidification microstructure is well documented in the literature, where matched asymptotic analysis is used to map PF models onto quantitative sharp interface theories that are quantified by a few a-priori known microscopic parameters \cite{Echebarria04,Plapp2011,Nikbook}. However, traditional PF theories lack any atomic-scale structure, thus precluding a explicit connection polycrystalline solidification, grain boundary energy, interface anisotropy, void formation and vacancy trapping, among others. Also, adding elasticity and defect flow is challenging, requiring the introduction of multiple added fields and assumed couplings between them. 

Classical density functional theory (CDFT) provides an alternate route for studying crystallization and solid state transformations  \cite{RAMAKRISHNAN79}. CDFT models are formulated in terms of a coarse grained mass density and employ multi-point correlations in the excess free energy to model  interactions that govern the properties of solid phases. This makes it possible to model a wide range of metallic and non-metallic materials.  CDFT theories also naturally give rise to grain boundaries and interface kinetics arising from the atomic structure of the interface.  The recent offshoot of CDFT, {\em phase field crystal (PFC) theory}, is a type of phase field theory with an atomic-scale order parameter related to the atomic mass density \cite{elder2004,elder2007}. PFC models are rotationally invariant and do away with the  need of multiple fields in order to model different crystal orientations, elastic fields or dislocations. In this paradigm grain boundaries and anisotropies are naturally captured \cite{Wu07,Sami2009,Wu10b} within the purview of a single order parameter field. These models also naturally (and with relative computational ease) give rise to a wide range of defect phenomena that are relevant to metals \cite{berry2011,berry2014}.  When coupled with noise, PFC models have also been used to elucidate nucleation in rapid solidification, as well as pathways for the formation of metastable phases \cite{granasy2011,emmerich2012,FALLAH13_AlCu_experiment,paulpaper}. 

An important aspect relevant to rapid solidification that has been lacking in previous PFC studies is the incorporation of thermal transport alongside the diffusive dissipative  dynamics of the PFC density order parameter. At slow rates of solidification, this is a negligible effect. Even at moderate rapid rates of solidification relevant to some types of laser welding of aluminum alloys (interface speed $v\sim 10^{-3}$m/s and thermal gradient $G\sim 10^6$K/m) this is negligible to lowest order as the scale of microstructure (typically $\lambda \sim 10\mu$m) is much smaller than the thermal diffusion length ($l_T =\alpha/v \sim 1$cm), and thermal diffusion time ($t_D = \alpha/v^2 \sim 1$s) is much greater than the $\sim 10^{-2}s$ solidification time of a $50\mu m$ powder. At the other extreme of rapid solidification rates, such as those used in single-pulse laser melting (interface speeds as high as $\sim 100 m/s$ and $G\sim 10^6K/m$), the thermal diffusion length is around  $l_T \sim 1 \mu$m, which is the same scale of microstructure, and the thermal diffusion time $t_D \sim 10^{-8}s$, about 
$\sim 2$ orders of magnitude faster than the $\sim 10^{-6}s$ solidification time of a $50\mu m$ powder. This implies that the thermal transport associated with latent heat of solidification is important to incorporate in a microscopic model of solidification. This is particularly crucial in determining nucleation undercooling conditions  that are so critical to rapid cooling of small metal powders. Another interesting physical phenomenon of interest is rapid crystallization of amorphous thin films. Here it has been conjectured that the delicate balance of latent heat release and thermal dissipation drives the  amorphous to crystalline transformation \cite{Wickersham78,explosive}.   

This paper derives a self-consistent formalism for coupling  PFC models of a pure material to temperature transport and thermal fluctuations. We start in Section~\ref{fluxes} by coupling the thermodynamic driving forces for mass and energy to their corresponding conservation laws, thus deriving a two  transport equations that couple thermal transport to microscopic mass density. This formalism is then specialized to two models in Section~\ref{cDFT_coupling}. The first is the classical CDFT model of Ref.~\cite{RAMAKRISHNAN79}. The second is the recent PFC model of Kocher, et al \cite{Kocher2015},  arriving at a model that couples the PFC order parameter to an equation for the effective PFC temperature scale. Section~\ref{apps1} examines the properties of the latter thermal-PFC model,  demonstrating thermal diffusion and latent heat release in 1D, as well as early stage nucleation, kinetic undercooling and recalescence in rapid solidification of 2D films.  Section~\ref{newmodel} introduces a new PFC type theory of a single component material that generalizes the work of Kocher, et al \cite{Kocher2015}; this decomposes the free energy into a modified Van der Waals contribution that controls the long wavelength properties of phases, and two excess contributions that, respectively,  control crystallographic and other short-range properties of a crystallizing system. The aim of this model is to unify previous PFC models of solidification under a more general PFC type model that can consistently represent the properties of pure materials. Section~\ref{newmodel} demonstrates the equilibrium properties of the new unified model, and some of its approximations. Section~\ref{UniHeat} ends by deriving the coupling of the unified model to thermal  transport by applying the formalism of Section~\ref{cDFT_coupling}.

\section{Thermodynamic Fluxes and Conservation Laws}
\label{fluxes}

Most formalisms of non-equilibrium thermodynamics start by relating changes in entropy to the fundamental thermodynamic fields of an evolving system. One manifestation is
\begin{equation}
ds = \frac{1}{T} d e - \frac{\mu}{T} d \rho
\label{first_law}
\end{equation}
where $s$ is the entropy density of a volume element, $T$ is temperature, $e$ is the internal energy density, $\rho$ is the local average mass density and $\mu$ is a chemical potential.
Eq.~(\ref{first_law}) can be taken to imply, in the mean field sense, that a volume element in a system is in local equilibrium, although it can vary from volume to volume such as to allow spatial variations of the relevant fields. We perform a functional generalization of eq~(\ref{first_law}) to model the spatial variation of the fields, giving
%
\begin{equation}
\delta S[e, \rho]=\int_V d\vec{x}^3 \,\, \frac{1}{T(x,t)}\ \delta e 
+\int_V d\vec{x}^3 \,\, \frac{1}{T(x,t)}\frac{\delta F[T,\rho]}{\delta \rho} \delta \rho 
\label{first_law2}
\end{equation}
where $F[T,\rho]$ is the free energy of the system, which depends on the temperature ($T$), density ($\rho$) and its gradients. 
As conserved fields, internal energy and density satisfy the conservation laws
\begin{eqnarray}
\frac{\partial e}{\partial t} &=& - \nabla \cdot \vec{J}_e \left( \frac{\delta S}{\delta e}, \frac{\delta S}{\delta \rho} \right) \nonumber \\
\frac{\partial \rho}{\partial t} &=& - \nabla \cdot \vec{J}_{\rho} \left(\frac{\delta S}{\delta e}, \frac{\delta S}{\delta \rho} \right)  
\label{cons1}
\end{eqnarray}
where $\vec{J}_e$ and $\vec{J}_\rho$ are, respectively, the energy and density fluxes. We postulate that these are linear functions of the thermodynamic driving forces,  $  \nabla \left[ {1}/{T(x,t)}\right]$ and $\nabla \left[{1}/{T(x,t)}\cdot{\delta F[T,\rho]}/{\delta \rho}\right]$. We derive at a self-consistent form of this theory using experiments and symmetry to guide us where appropriate. 

As a minimal description, we start by assuming that the cross effects in mass and energy diffusion can be neglected. We also choose the phenomenological Onsager coefficient for energy diffusion as $L_{uu}=KT^2/2$, where $K$ is the thermal conductivity. This form is chosen to recover Fick's law,
\begin{equation}
J_e=\frac 1 2 KT^2 \nabla \frac 1 T =-K \nabla T
\end{equation}
Writing the Onsager coefficient for density diffusion as $\Gamma \rho$, where $\Gamma$ may depend on temperature, gives the classic CDFT density flux for $J_\rho$.  Combining these fluxes we can now write two conservation equations (\Es{cons1}) for energy and mass transport as
\begin{eqnarray}
\frac{\partial e}{\partial t} &=& K\nabla ^2 T \nonumber \\
\frac{\partial \rho}{\partial t} &=& \nabla\left(\Gamma(T) \rho\nabla \left(\frac 1 T \frac{\delta F}{\delta \rho}\right)\right)
\label{cons2}
\end{eqnarray}
To arrive at a closed form model with thermal coupling, we must specify the free energy and its temperature and density dependences. The details of this process for the PFC model are reserved for the next section. Here, we just mention that in general, an expression for the energy density $e$ can be derived from the free energy density by using the functional generalization of $s=-\left. {d f}/{d T} \right|_{\rho}$ that allows for gradient dependence in the free energy density, i.e.,
\begin{equation}
e= f(T, \rho, \nabla \rho)-T \frac{\delta F[T, \rho, \nabla \rho]}{\delta T}
\label{free-to-e2}
\end{equation}

Note that the density equation in Eq.~(\ref{cons2}) is analogous to the one used in Density Functional Theory (DFT). In its original derivation \cite{DDFT}, $\rho$ was interpreted as an ensemble averaged quantity, which is why formally no noise is added to this equation. When applied to PFC modelling (or phase field modelling), noise will be added to account for stochastic events like nucleation.
\section{Temperature coupling for the basic cDFT and PFC models}
\label{cDFT_coupling}
This section specializes the formalism of the last section first to the classical CDFT model of Ref.~\cite{RAMAKRISHNAN79}, and then to the PFC model of Kocher, et al \cite{Kocher2015}.

\subsection{Application to CDFT theory of freezing}
We  consider the general free energy functional of a simple CDFT type theory of the form
\begin{eqnarray}
F&=& \int d\bx \,f  \label{pfcfree}\\
&=&\int d\bx \, k_BT\left\{ \rho \ln\frac {\rho}{ \bar{\rho}}    \!-\! \delta\rho - \frac 1 2 \int d \bx'\delta\rho(\bxp) C_2\delta\rho(\bx)\right\} \nonumber 
\end{eqnarray}
where $C_2$ is the two point density-density correlation function of the theory. It is nominally taken at the reference density $\bar{\rho}$ of the liquid at coexistence, but we tacitly assume that the correlation function has some $T$ dependence away from the reference. 
The free energy density $f$ varies on the length scale of the density $\rho$. Temperature, however, varies much more slowly than density.  As a result, a sensible energy conservation equation of the form appearing in \E{cons2}  should consider $e$ coarse grained on the same length scale as the variation in the temperature. This is done by applying a smoothing operator $\chi$ on the microscopic internal energy $e$, or any quantity that involves the microscopic free energy functional $f$. 

With the above considerations, we begin by computing the right hand side of Eq.~(\ref{free-to-e2}) using Eq.~(\ref{pfcfree}), which gives, 
\begin{equation}
\frac{\partial F}{\partial T}\!=\! f/T \!+\! \int dx'\left[ -\frac 1 2 k_BT \delta \rho(x) C_2^{\, \prime}(x,x',T) \delta \rho(x')  \right]
\label{dfdt}
\end{equation}
where the notation $()'=\partial () /\partial T$ is introduced for functions of $T$. Substituting \E{dfdt} into  \E{free-to-e2} and coarse graining gives 
\begin{eqnarray}
e&=&\chi\!*\!\left[f-T  \frac {\delta F}{\delta T  } \right] \nonumber \\
&=& \frac 1 2 k_BT^2 \chi\!*\!\left[ \int dx'\left[ \delta \rho(x)C_2'(x,x',T)\delta\rho(x') \right] \right]
\label{e_form_PFC}
\end{eqnarray}
where $*$ denotes the convolution operation with the smoothing function $\chi$. Eq.~(\ref{e_form_PFC})  is a suitable form for internal energy to use in \E{cons2}.  Note that we have pulled the smoothing operation through the field $T$ because we assume temperature is a smooth variable on the scale of atomic variations inherent in the PFC density $\rho$. The time derivative of \E{e_form_PFC} gives
\begin{eqnarray}
\dot{e}&=&k_B\dot{T}\chi\!*\!\left[\int dx' \delta\rho(x)\left( TC_2'+\frac 1 2 T^2C_2'' \right)\delta\rho(x') \right]\nonumber\\
	  &+&k_BT\chi\!*\!\left[ \int dx' \delta\dot{\rho}(x) TC_2' \delta\rho(x') \right] \label{e_dott},
\end{eqnarray}
substituting \E{e_dott} into the first of eq.~(\ref{cons2}) gives a heat equation for the general CDFT model, 
\begin{eqnarray}
&& \chi\!*\!\left[\int dx' \delta\rho(x)\left( TC_2'+\frac 1 2 T^2C_2'' \right)\delta\rho(x') \right]\, \frac{\partial T}{\partial t} =\nonumber \\ 
&& \frac{K}{k_B}\nabla^2T \nonumber - \chi\!*\!\left[ \int dx' \delta\dot{\rho}(x) TC_2' \delta\rho(x') \right] T
\label{tempeq1}
\end{eqnarray} 

\subsection{Specialization to the Vapor-PFC Model}
\label{pfcheatcoupling}

We next proceed to specialize the above general PFC temperature equation to the recent Vapor PFC model derived in \cite{Kocher2015}. The vapour PFC model was illustrated using the same two point correlation  of the original PFC model, 
\be
\bar{\rho}C_2(\bx,\bx',T)=\left[1-r-B_x\left(1+R^2 \nabla_x^2\right)^2\right]\delta(\bx-\bx'), \label{pfcC2}
\ee
where  $R$ tunes the lattice constant of the PFC solid phase and $\nabla_x$ refers to differentiation with respect to dimensional variables. Here, $B_x$ and $r$ are dimensionless constants. The parameter $r$ is the effective temperature parameter of the original PFC model
\footnote{The coefficients $r$, $B_x$ and $R$ are formally related to the coefficients $\hat{C}_n$ of the Fourier space expansion of $ \bar{\rho}  C(|\vec{x} - \vec{x'}|)=  \left( -\bar{\rho} \hat{C}_0 - \bar{\rho} \hat{C}_2 \nabla^2 - \bar{\rho} \hat{C}_4 \nabla^4 \right) \delta (\vec{x} -\vec{x'} )$, which are evaluated at some reference density and reference temperature $(\bar{\rho}, T_0)$ on the liquidus of the phase diagram. In the PFC literature, it is tacitly assumed that as one moves away form $(\bar{\rho}, T_0)$ in temperature, these coefficients can vary with temperature in some sensible way. It is in this way that $r$ is typically assigned the role of PFC temperature; $B_x$ and $R$ are typically considered be constant.}.
 The addition of the new 3rd and 4th order low-k mode correlation terms introduced in \cite{Kocher2015} yields the following free energy functional,
\bea
 F&=&\int d\bx \, k_BT \bar{\rho}  \left\{n\left(r +B_x(1+ R^2 \nabla_x^2)^2\right)\frac{n}{2}\right.\nonumber\\
 &-&\left.\frac{n^3}{6}+\frac{n^4}{12} + \left( a \frac{\bar{n}^2}{3}+b\frac{\bar{n}^3}{4}\right) n \right\},
 \label{pfc_detailed}
\eea
where $a$ and $b$ are constants, and where we have transformed to the dimensionless density $n(\bx,t)=(\rho(\bx,t)-\bar{\rho})/\bar{\rho}$, while 
\be
\bar{n}=\chi*n\equiv\int d\bx' \chi(\bx-\bx')n(\bx)
\ee
It is noted that in this minimal model the effective 3 and 4-point terms do not depend on temperature. This form will serve to illustrate most of the physical features of density-temperature coupling, although it isn't as robust as if were to assume that these terms also depend on the temperature  (more on this in Section~\ref{newmodel}). In this derivation of a heat equation coupled to the PFC equation, we assume that only the PFC temperature scale $r$ depends on physical temperature. One may additionally assume a temperature dependence for $R$, but we neglect this effect here.

To proceed with the first of \E{cons2}, we start by  scaling the right hand side. We first assume a mapping from $T$ to $r$ of the form $T(\bx,t)=T_0\theta(r(\bx,t))$, where the dependency on $x$ is written here to emphasize the spatial variation of $r$. It is noted that like $T$, the PFC temperature $r$ is smooth at the atomic scale. This gives 
\be
K \nabla^2 T=KT_0\nabla_x^2\theta(r)= \frac{KT_0}{R^2}\, \nabla^2\theta(r)
\label{heat_eq2}
\ee
where the last expression in \E{heat_eq2} assumes length is rescaled  according to $\br=\bx/R$, and $\nabla$ denotes dimensionless derivatives with respect to $\br$. Using the chain rule gives
\be
\nabla^2\theta(r)=\nabla\cdot(\nabla \theta(r))=\nabla\cdot(\theta'\cdot\nabla r)=\theta'\nabla^2r+\nabla \theta'\cdot\nabla r
\label{k_gradients}
\ee
Substituting these gradients back into the internal energy equation of \E{cons2} gives
\bea
\frac{\dot{e}}{\bar{\rho}k_BT_0}&=& \frac{K}{\bar{\rho} R^2k_B}(\theta'\nabla^2r+\nabla \theta'\cdot\nabla r) \nonumber \\
&=&\mathcal{C}^{*}(\theta'\nabla^2r+\nabla \theta'\cdot\nabla r)
\label{rhs}
\eea
where $C^*=K/(\bar{\rho} R^{2}k_B)$ has units of $s^{-1}$. Next, we coarse grain the local internal energy density, taking \E{pfc_detailed} as input, i.e.,
\begin{eqnarray}
e&=&\chi\!*\!\left(f-T\frac{\delta F}{\delta T(\vec{x},t) }\right)\nonumber\\
&=& -\chi\!*\!\left(\frac{\bar{\rho} k_B T^2 \, r^\prime }{2} n^2\right)
\label{e_eq}
\end{eqnarray}
This leads to 
\be
e/(\bar{\rho} k_B T_0^2)= -\chi\!*\!\left\{\theta(r)^2 r' n^2/2\right\}
\label{intermediate}
\ee
A formal expression for $r'$ can be found by differentiating
\be
\frac {\partial}{\partial T}T=T_0\frac {\partial}{\partial T}\theta\left(r(T)\right)=T_0\frac {\partial \theta}{\partial r}\frac{\partial r}{\partial T}=T_0 \theta' r'=1,
\ee
giving
\be
r'=\frac 1 {T_0\theta'}
\ee
Thus, \E{intermediate} becomes
\be
e/(\bar{\rho} k_B T_0)= -\theta^2/\theta' \chi\!*\!\left\{n^2/2\right\},
\label{intermediate2}
\ee
where functions of $\theta(r)$ were taken out of the smoothing operations as they vary on long wavelengths by hypothesis. Taking the time derivative of \E{intermediate2} gives
\bea
\dot{e}/(\bar{\rho} k_B T_0)&=& \frac{(\theta^2\theta''-2\theta\theta'^2)}{2\theta'^2} \chi\!*\!\left[n^2/2\right]\frac{\partial r}{\partial t} \nonumber\\
&-&\theta^2/\theta' \, \frac{\partial }{\partial t}\left\{\chi\!*\!\left[n^2/2\right]\right\}
\label{e_dot}
\eea
Equating the right hand sides of \E{e_dot} and (\ref{rhs}) finally gives 
\bea
\frac{(\theta^2\theta''-2\theta\theta'^2)}{2\theta'^2} &\chi&\!*\!\left[n^2/2\right] \frac{\partial r}{\partial t}\nonumber\\
&=&\mathcal{C}^{*}(\theta'\nabla^2r+\nabla \theta'\cdot\nabla r)\nonumber\\
&+&\theta^2/\theta' \, \frac{\partial }{\partial t}\left\{ \chi\!*\!\left[n^2/2\right]\right\}
\label{PFC_heat_eq}
\eea

To study a  minimal model of the PFC heat equation, \E{PFC_heat_eq}, we neglect the $\nabla \theta'\cdot\nabla r$ term in \E{PFC_heat_eq}. Also, one must determine  the form of $\theta(r)$. We assume that $\theta(r)$ needs to satisfy $\theta'>0$ ($r$ is an increasing function of temperature) and we also assume that $F(T)$ is a concave functional of temperature $\delta^2 F/\delta T^2<0$ for thermodynamic stability. To see the constraint that this condition sets, we calculate
\be
\delta^2 F/\delta T^2=\bar{\rho}k_B\frac{n^2}{2}(2r'+Tr'')
\ee
Since $r'=1/(T_0\theta')$, $r''=-1/T_0^2\theta''/\theta'^3$, the condition $\delta^2 F/\delta T^2<0$  amounts to $(2\theta'^2-\theta\theta'')<0$, or just $(\theta^2\theta''-2\theta\theta'^2)>0$, assuming $\theta>0$. This is exactly the term found on the left hand side of \E{PFC_heat_eq}, and confirms that the diffusion and source pre-factors remain positive given the right $\theta$ function. 

The lowest order $\theta(r)$ function that satisfies the two constraints is a quadratic function. For different materials, different fitting functions can be found to match $T$ to $r$. To proceed here we will take the factors that involve $\theta(r)$, $\theta'(r)$ and $\theta''(r)$ in \E{PFC_heat_eq} as constants  as we are not particularly interested in the quantitative details of how temperature affects them; varying  these parameters  changed results by 5-10\% over the range of model temperature range simulated, and their form is only of quantitative interest once such a model is applied to a specific material. We therefore simplify the \E{PFC_heat_eq} to a minimal  heat equation for the PFC model given by 
\be
{ \chi\!*\!\left[n^2/2\right]} \frac{\partial r}{\partial t}={\mathcal{C}_d\nabla^2r+C_s\frac{\partial }{\partial t}\left\{\chi\!*\!\left[n^2/2\right]\right\}},
\label{finalmodel}
\ee
where $\mathcal{C}_d$ (units $s^{-1}$) and $\mathcal{C}_s$ (dimensionless) are parameters of the theory. It is noteworthy that \EE{finalmodel} has term accounting for latent heat release and a susceptibility term, both directly linked to changes in PFC density. 

\section{Simulations of heat transfer in the PFC formalism}
\label{apps1}

The objective of this section is to demonstrate the  consistency of \E{finalmodel} when used with free energy $F$ given bu \E{pfc_detailed}. We couple heat transfer to the standard PFC density dynamics \cite{elder2002},
\begin{equation}
\partial_t{n(\br)}=\Gamma \nabla^2\frac{\delta \mathcal{F}[ n(\br,t)]}{\delta n(\br,t)}+\eta(\br),
\label{PFC_dyn}
\end{equation}
where $\mathcal{F}=F/(k_BT_0 \bar{\rho} R^d)$ and space has been rescaled according to $\br=\bx/R$. The term $\eta(\br)$ is a stochastic noise term used to model thermal fluctuations. It follows a Gaussian statistics  with amplitude  scaled by factor ${N}_a$, and is wavelength-filtered as prescribed in Ref.~\cite{Kocher2016} to assure that interface fluctuations are consistent with capillary fluctuation theory.  In these units, $[\Gamma]=s^{-1}$, and so we will define a characteristic time  $\bar{t}=1/\Gamma$, and scale time in our equations as $t\rightarrow t/\bar{t}$. This amounts to setting $\Gamma=1$ in \E{PFC_dyn}. Also for simplicity, we will demonstrate \E{finalmodel} here for the free energy of the standard PFC model, hence setting the $a$, $b$ coefficients of the  vapour PFC model to zero, i.e. $a=b=c=0$. The PFC equation in \E{PFC_dyn} is coupled to heat equation  
\be
{ \chi\!*\!\left[n^2/2\right]} \frac{\partial r}{\partial t}={\mathcal{C}_d\nabla^2r+C_s\frac{\partial }{\partial t}\left\{\chi\!*\!\left[n^2/2\right]\right\}} \!+\!\mathcal{C}_b,
\label{simple_T_eq}
\ee
which is the same as \E{finalmodel} except with an added term $\mathcal{C}_b$ to account for heat extraction  out of the simulation domain. In In the remainder of this section, we promote $\mathcal{C}_d$, $\mathcal{C}_s$ and $\mathcal{C}_b$ to be simple constants of the model. We will return to the question of a more quantitative PFC theory, and its corresponding couplig to heat transfer int the last two sections.

\begin{figure}[h!]
\centering
\includegraphics[width=0.5\textwidth]{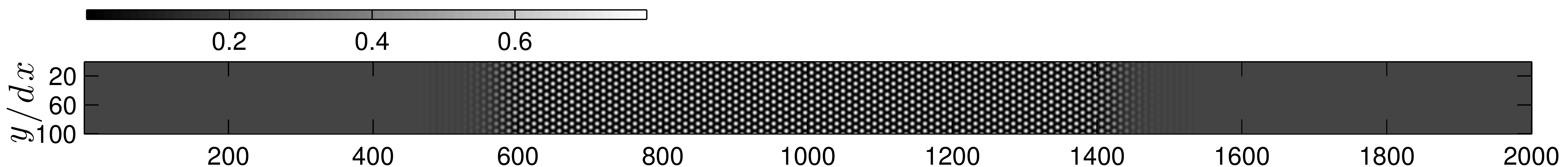}
\includegraphics[width=0.5\textwidth]{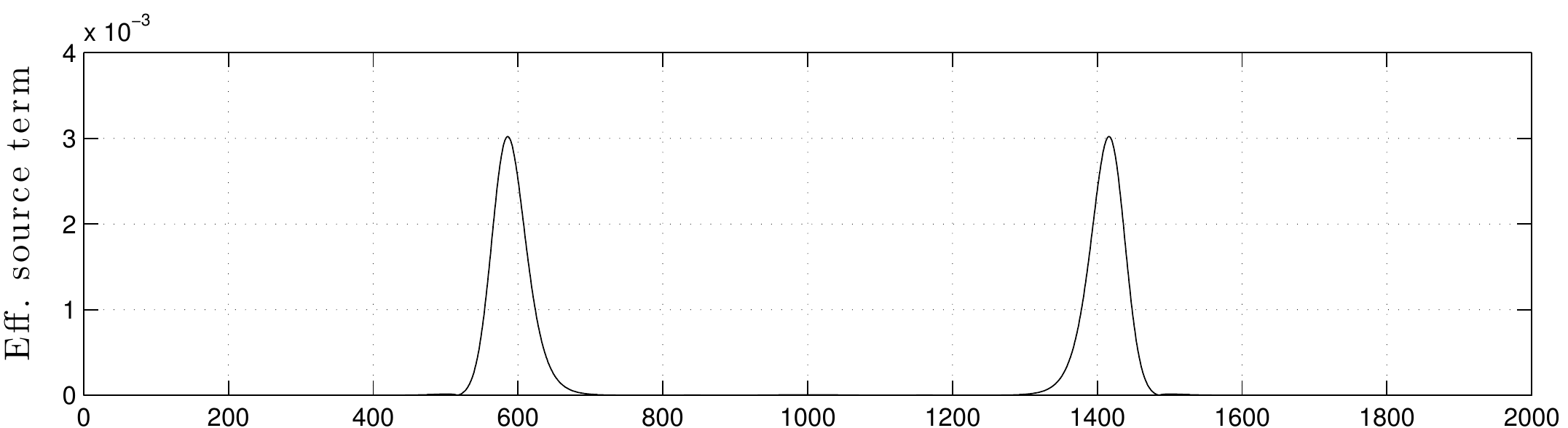}
\includegraphics[width=0.5\textwidth]{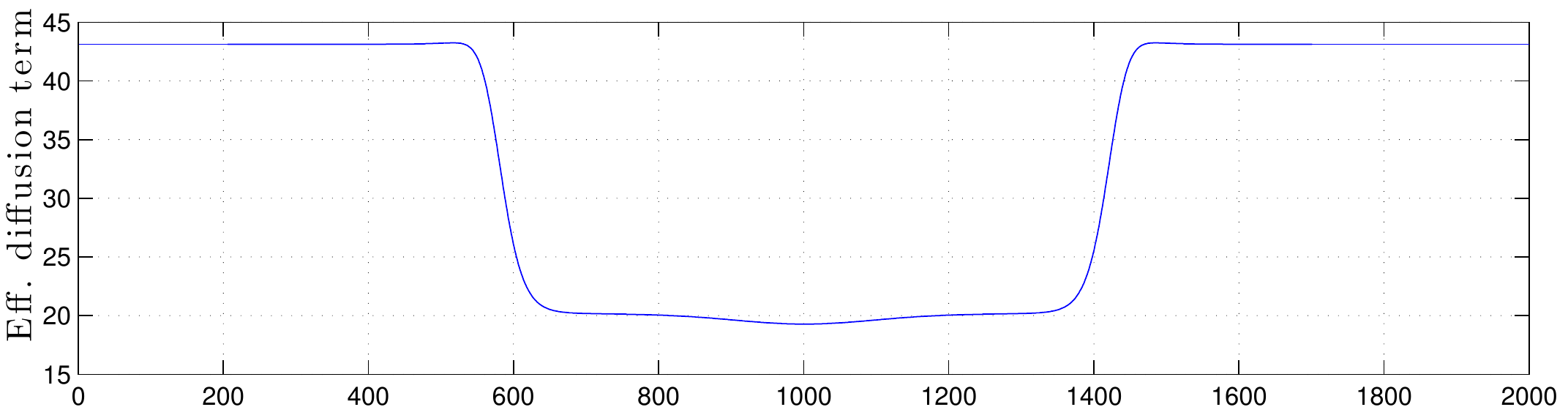}
\includegraphics[width=0.5\textwidth]{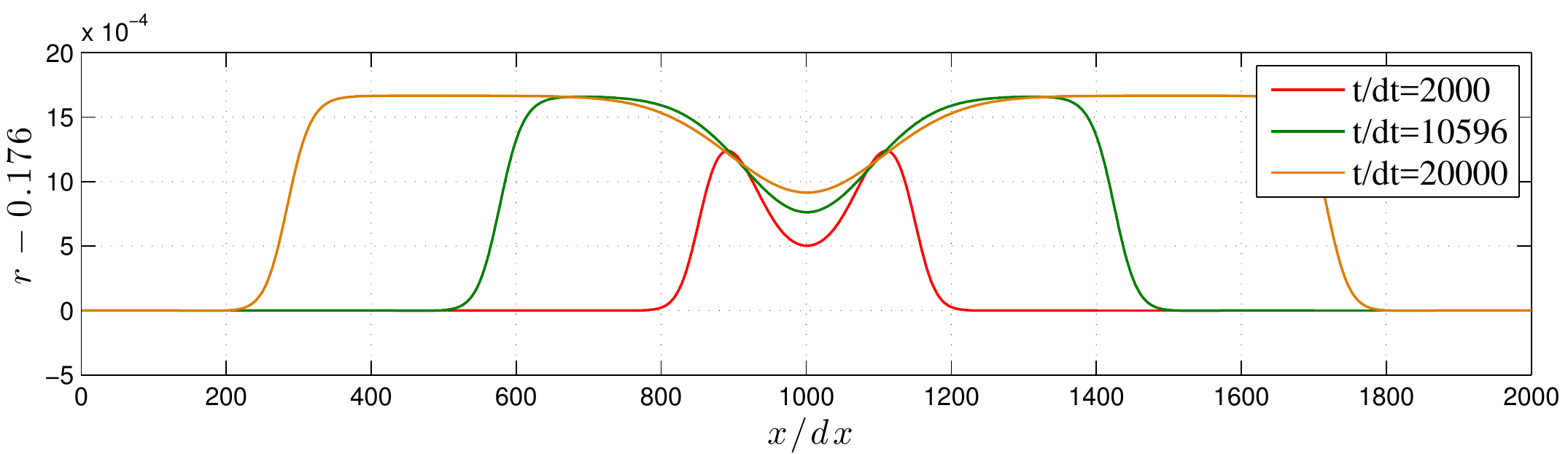}
\caption[Growth of a solid slab in the thermal PFC model]{Growth of a solid slab in the PFC model coupled to thermal transport. The top panel shows the density field after 10596 numerical time steps, while the other three show plots of local quantities at $y/dx=50$ as a function of $x$. Top middle is the source term $\partial_t\! \left\{\chi*\left[n^2/2\right]\right\}/{ \chi\!*\!\left[n^2/2\right]}$. Bottom middle is the effective diffusion coefficient $1/{ \chi\!*\!\left[n^2/2\right]}$, while the bottom plot is the temperature at different times. Model parameter values: $B_x=1.5$, quench temperature  $r_0=0.176$, average density $\langle n\rangle\equiv n_0=0.21$, $\mathcal{C}_s=0.001$, $\mathcal{C}_d=0.02$, $\mathcal{C}_b=0$, noise amplitude $N_a=0$, $dt=0.25$, and $dx=0.7256$. \label{fig:slab_thermal}}
\end{figure}

We demonstrate our formalism by considering a simulation of a small 2D solidifying slab. As a first test, we consider the case where 
$\mathcal{C}_b=0$ and shut off the microscopic fluctuations  in the PFC density equation. A quenched liquid phase is brought in contact with a solid slab of lateral width $100 dx$ (out of a total $2000 dx$ in the x direction). After initial equilibration, the slab of solid grows, consuming the liquid (see top frame in \ref{fig:slab_thermal}). Due to the rearranging density, a non zero source term is generated in the temperature equation by the term $\partial_t \left\{\chi*\left[n^2/2\right]\right\}$. This term is peaked around the regions of high density changes, i.e. the moving interface in this case.  This leads to a temperature increase in and behind the interface. Some profiles are shown in the bottom frame in 
Fig.~\ref{fig:slab_thermal}, for early, intermediate and late time. Since $\mathcal{C}_b=0$, the heat generated remains inside the system, and, after the interface advances through the system, the temperature reaches a plateau higher than the starting temperature of the liquid. The model's effective temperature diffusion term is scaled by $1/{( \chi\!*\!\left[n^2/2\right])}$, which is spatially dependent. This term is shown in the bottom-middle frame in Fig.~\ref{fig:slab_thermal}, and is seen to take on very different values in the liquid and in the solid. It also reacts to smaller density and amplitude differences inside the solid phase, as seen by the slight dip in the center. The top middle frame in Fig.~\ref{fig:slab_thermal} shows the last term in \E{simple_T_eq}. which accounts for latent heat.

\begin{figure}[h!]
\centering
\includegraphics[width=0.5 \textwidth]{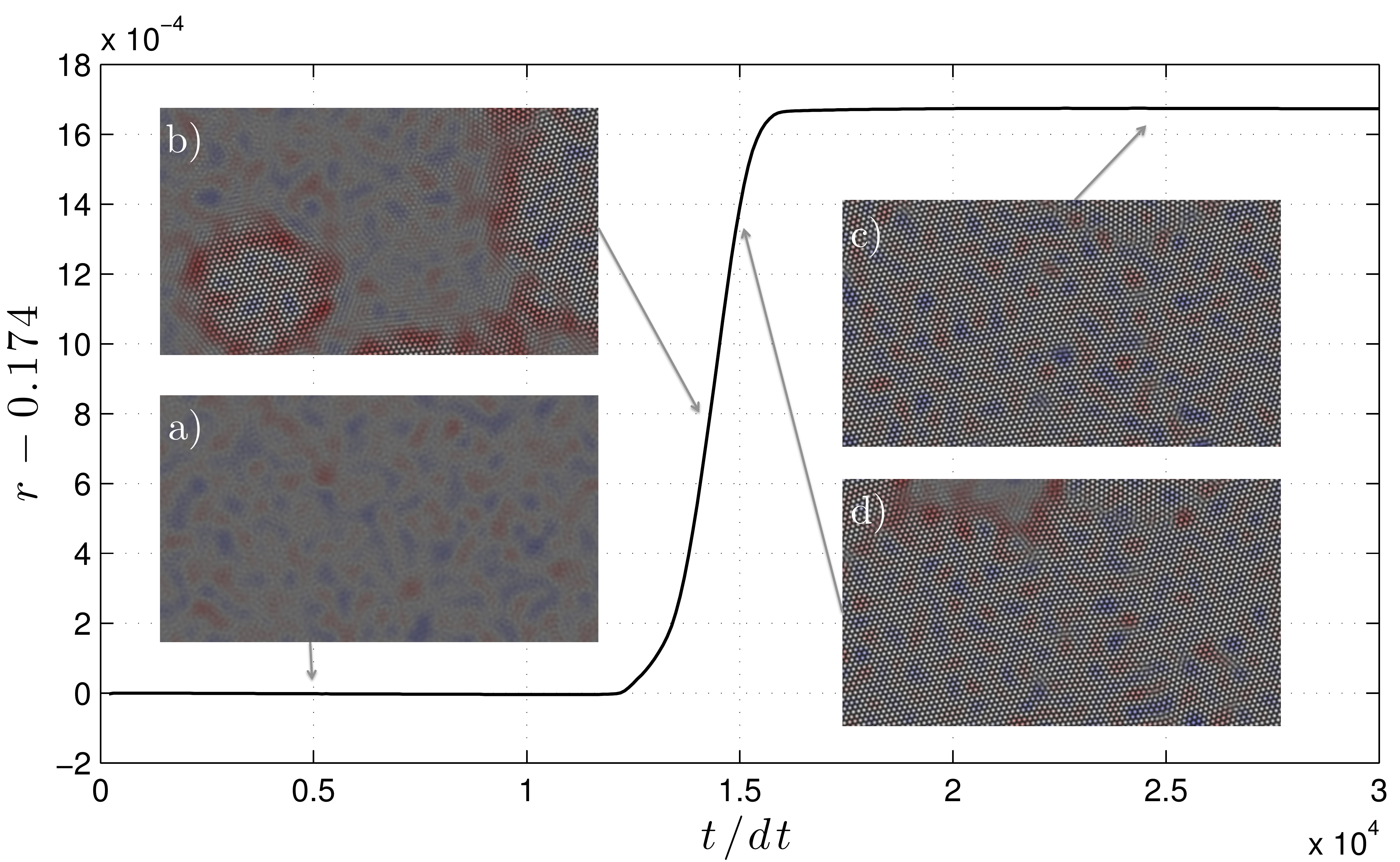}
\caption[Nucleation in the Thermal PFC model]{The average temperature versus time, during a nucleation simulation shown in insets a) to d), which correspond, respectively, to $5000 dt, 14000 dt, 15000 dt$ and $25000 dt$. The density field is coloured to show the magnitude of the temperature source term, where red is a positive source term while blue is a negative source term.  No colouring corresponds to nearly zero source term. Parameter values: $B_x=1.5$, $r_0=0.174$, $n_0=0.225$, $\mathcal{C}_s=0.001$, $\mathcal{C}_d=0.005$, $N_a=0.04$, $dt=0.5$, $\lambda=0.1$. \label{fig:recalescence_test}}
\end{figure}

We tested the effect of nucleation on temperature in an adiabatic system. We ran a simulation of a supersaturated liquid, perturbed by density fluctuations. While the full simulation was performed in a $2000 dx \times 2000 dx$ box, only a small portion of this domain is shown in the insets of Fig.~\ref{fig:recalescence_test}. As expected physically, several seeds eventually nucleate and grow, while releasing latent heat at interfaces. This can be seen in the insets of Fig.~\ref{fig:recalescence_test}, where the thermal source term is overlaid as a colour map on top of the PFC density field. This is also reflected in the main part of Fig.~\ref{fig:recalescence_test}, which shows the average system temperature versus time; since the system is adiabatic, the average temperature of the system increases.

\begin{figure}[h!]
\centering
\includegraphics[width=0.48\textwidth]{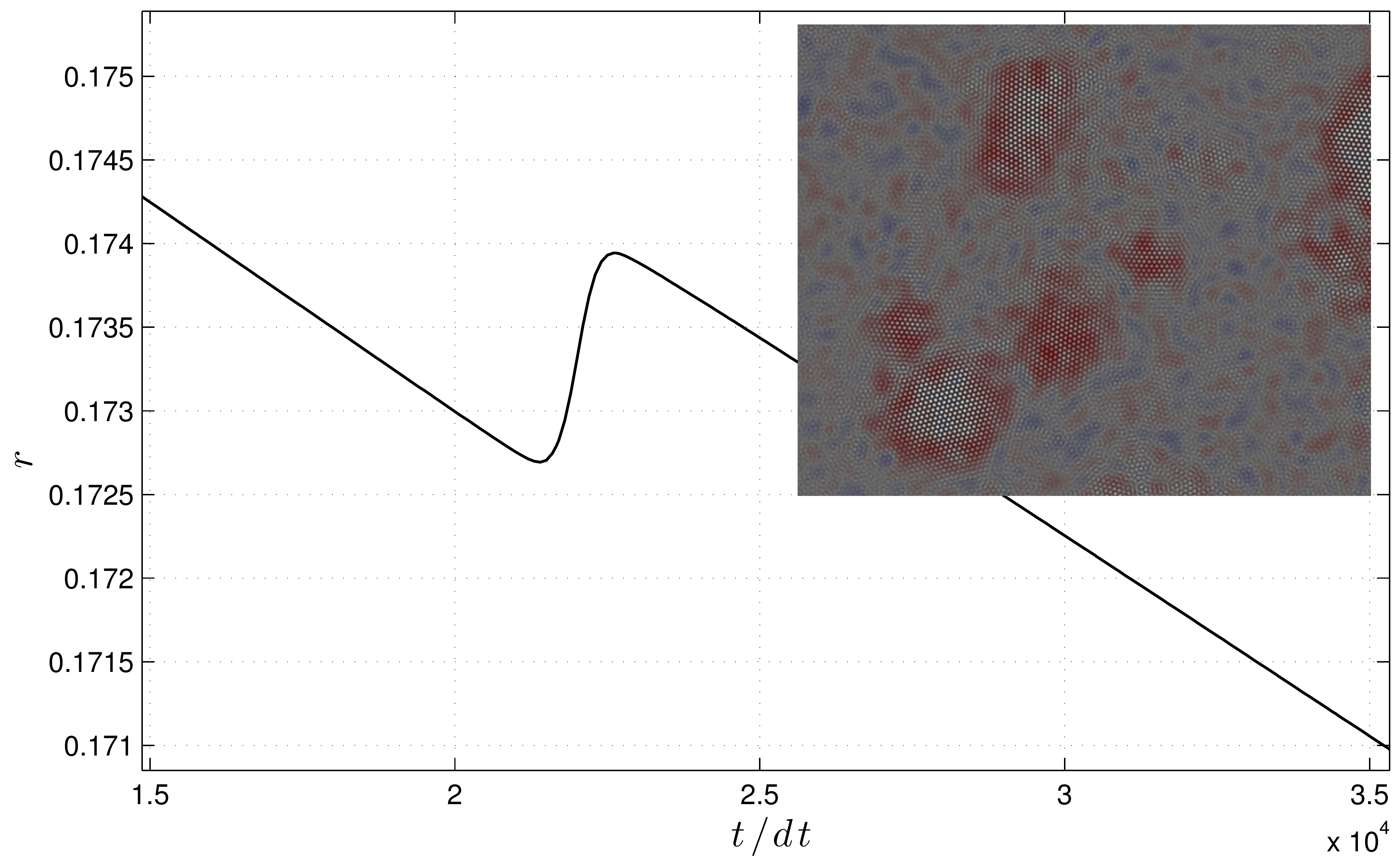}
\caption[Nucleation in the Thermal PFC model.]{Nucleation and recalescence in the thermal PFC model, with constant heat extraction of $\mathcal{C}_b=10^{-6}$. The figure shows the average system temperature as a function of time, while the inset show a crop of the density field at $t/dt=21661$. The colour map shows the intensity of the source term in the temperature equation. Parameter values are the same as Fig.~\ref{fig:recalescence_test}.
\label{fig:recalescence_nucleation} }
\end{figure}

As a more stringent test of the robustness of our density-temperature PFC formulation, we proceed with a nucleation simulation in a constantly cooled system in a $2000 dx \times 200 dx$ domain. A supersaturated liquid system is initialized at a high temperature (in the solid region to expedite the simulation time), and uniform heat extraction is modelled by activating the $\mathcal{C}_b$ term in \E{simple_T_eq}.  Density fluctuations are once again incorporated. The resulting average temperature field is shown in Fig.~\ref{fig:recalescence_nucleation}, and the inset shows the corresponding density field during the nucleation process. The figure shows a strong temperature spike during nucleation, followed by a decrease in temperature as the system continues to cool.

The above  simulations show the consistency of our approach for coupling the PFC density order parameter to thermal transport through the PFC model's effective temperature field $r({\bf x},t)$. We have demonstrated that in the absence of noise, behaviour very close to those of more traditional model C type dynamics are observed, with an effective diffusion coefficient and latent heat source that couple directly to the microscopic density field. We have also shown that the formulation is robust to density noise. Experimentally relevant thermal traces of microstructure rearrangements, like recalescence for example, can be observed using this technique.
\subsection{Latent heat: mesoscopic limit of thermal PFC model}

One of the key features of \E{finalmodel} (or \E{PFC_heat_eq}) is that latent heat is self consistently coupled to the PFC density. It is instructive to derive an expression for the latent heat for the new PFC thermal model by studying its long wavelength, or phase field, limit analogously to the  analyses done  \cite{Nikbook} on {\it model C} \cite{Hoh77}. 

We first consider a planar solidification front. We start by re-writing eq.~ (\ref{finalmodel}) in the co-moving reference frame by making the substition $\partial_t\rightarrow\partial_t-v\partial_x$, where $v$ is the normal velocity of the front. This gives
\be
\partial_t r-v\partial_xr=2\frac{C_d\partial_x^2r+C_s(\partial_t-v\partial_x)\chi*\left[n^2\right]}{\chi*n^2}
\ee
Assuming a stationary situation, all time varying terms can be neglected giving
\be
-v\partial_xr =2C_d\frac{\partial_x^2r}{\chi*n^2}-2vC_s\frac{\partial_x\left(\chi*\left[n^2\right]\right)}{\chi*n^2}
\label{calculation}
\ee
We now consider a 1-mode approximation of the density,
\be
n(\br)=n_0(\br)+\sum_\textbf{G} \phi(\br) e^{i \textbf{G}\cdot\br} +c.c.,
\ee
where $n_o(\br)$ and $\phi(\br)$ are assumed to vary on length scales much larger than the atomic scale
implied by $2\pi/|\textbf{G}|$. Substituting this form into $\chi * n^2$ and coarse graining using standard box-averaging techniques \cite{Sami2009} gives 
\be
\chi*n^2=n_0^2+\nu \,|\phi|^2,
\ee
where $\nu=2$ in 1D and $\nu=6$ in 2D. Substituting the above coarse grained expression into \E{calculation} gives

\be
-v\partial_x r =2C_d\frac{\partial_x^2r}{n_0^2+\nu|\phi|^2}-2vC_s\frac{\partial_x\left(n_0^2+\nu|\phi|^2\right)}{n_0^2+\nu|\phi|^2}
\label{calculation2}
\ee

It is reasonable to assume that the temperature field is much smoother than both the average density and the amplitude, which vary on the scale of the solid-liquid interface width $2 \epsilon$ ($\sim$ a few nm in metals). Assuming the interface is centred on $0$, we integrate across the interval from $-\epsilon$ to $+\epsilon$ yielding
\bea
-v\int_{-\epsilon}^{+\epsilon}dx \, \partial_xr &=&
2C_d\int_{-\epsilon}^{+\epsilon}dx\frac{\partial_x^2r}{n_0^2+\nu|\phi|^2} \nonumber\\
&-& 2vC_s\int_{-\epsilon}^{+\epsilon}dx\frac{\partial_x\left(n_0^2+\nu|\phi|^2\right)}{n_0^2+\nu|\phi|^2}\nonumber
\eea
Completing the first and last integrals gives
\be
v[r]_{-\epsilon}^{+\epsilon} =
-2C_d \int_{-\epsilon}^{+\epsilon}dx\frac{\partial_x^2r}{n_0^2+\nu|\phi|^2}
+2v C_s\left[\ln\left(n_0^2+\nu|\phi|^2\right)\right]_{-\epsilon}^{+\epsilon}
\label{SIM_2}
\ee
In the limit of $\epsilon\rightarrow0$, $r(\br)$ remains a smooth function across the interface in comparison to $n_0(\br)$ and $\phi(\br)$, which become step functions to lowest order on the scale of variation of $r(\br)$. Taking $-\epsilon$ to be the solid side of the interface interval and $+\epsilon$  the liquid side, and integrating the remaining integral in \E{SIM_2} by parts gives, 
\bea
\!\! \int_{-\epsilon}^{+\epsilon}\!\!dx\frac{\partial_x^2r}{n_0^2+\nu|\phi|^2}\!&=&
 \!\frac 1 {n_l^2}\left. \frac{\partial r}{\partial x}\right|_{\epsilon^+} \!\!- \frac 1 {n_s^2+\nu|\phi_s |^2}\left. \frac{\partial r}{\partial x}\right|_{\epsilon^-} \!\! \nonumber\\
 &-& \int_{-\epsilon}^{\epsilon} \!dx\frac{\partial}{\partial x}\!\left(\frac{1}{n_o^2+\nu |\phi|^2}\right) \frac{\partial r}{\partial x},
 \label{SIM3}
\eea
where $\phi_s$ is the order parameter of the solid phase and $n_s$ and $n_l$ are the solid and liquid average densities, respectively. The  derivative in the last integral  in \E{SIM3} is a sharply peaked function at the interface on the scale of the smooth function $\partial r/\partial x$. The last integral in \E{SIM3} thus remains bounded and vanishes as $\epsilon \rightarrow 0$.  This can also be seen by integrating the last integral in \E{SIM3} by parts and approximating $r\approx const$ at $x=0$. This makes it possible to remove $r$ from the integral, leaving an odd function around $x=0$, which vanishes as $\epsilon\rightarrow 0$. Substituting \E{SIM3} into \E{SIM_2} thus yields, in the limit of small $\epsilon$,  
\bea
0&=& 2C_d\left(
\frac{1}{n_s^2+\nu|\phi|^2}\left. \frac{\partial r}{\partial x}\right|_{\epsilon^-} -\frac{1}{n_l^2}\left. \frac{\partial r}{\partial x}\right|_{\epsilon^+}\right)\nonumber\\
&-&2vC_s\text{ln}\left(\frac{n_s^2+\nu|\phi_s|^2}{n_l^2}\right),
\label{flux_consv1}
\eea
Rearranging the terms in \E{flux_consv1} gives 
\be 
 \frac {2C_d}{n_s^2 +\nu|\phi_s|^2}  \left. \frac{\partial r}{\partial x}\right|_{\epsilon^-}
\!-\frac {2C_d}{n_l^2}\left. \frac{\partial r}{\partial x}\right|_{\epsilon^+}
\!\!=2v\, C_s\ln\left(\frac{n_s^2+\nu|\phi_s|^2}{n_l^2}\right)
\label{flux_consv2}
\ee
\EE{flux_consv2} is the classic heat flux conservation across the interface. It will be shown below that the logarithmic term is proportional to the latent heat.

We can also analyze \E{calculation2} by integrating both sides from $-\infty$ to $+\infty$, analogous to projection operator approaches \cite{elder2001}. This gives
\bea
-v\int_{-\infty}^{+\infty}dx\partial_xr &=&2C_d \! \int_{-\infty}^{+\infty}dx\frac{\partial_x^2r}{n_0^2+\nu|\phi|^2}\nonumber\\
&-&2vC_s \!\int_{-\infty}^{+\infty}dx\frac{\partial_x\left(n_0^2+\nu|\phi|^2\right)}{n_0^2+\nu|\phi|^2}
\label{SIM3.5}
\eea
The middle integral in \E{SIM3.5} is zero in the limit of $\epsilon \rightarrow 0$. This can be shown by breaking the integral into three pieces as follows,  
\bea
\int_{-\infty}^{+\infty}\!\!dx\frac{\partial_x^2r}{n_0^2+\nu|\phi|^2}&=&
\int_{-\infty}^{-\epsilon}\!\!dx\frac{\partial_x^2r}{n_0^2+\nu|\phi|^2}\nonumber\\
&+&\int_{+\epsilon}^{+\infty}\!\!dx\frac{\partial_x^2r}{n_0^2+\nu|\phi|^2}\nonumber\\
&+&\int_{-\epsilon}^{+\epsilon}\!\!dx\frac{\partial_x^2r}{n_0^2+\nu|\phi|^2}  
 \label{SIM4}
\eea
Using the approximations that $n_o =n_s$ and $\phi=\phi_s$ for $x>\epsilon$ and $n_o=n_l$ and $\phi=0$ for $x<- \epsilon$ in the first two integrals on the right hand side, and using the results of \E{SIM3}, gives
\bea
\int_{-\infty}^{+\infty}\!\!dx\frac{\partial_x^2r}{n_0^2+\nu|\phi|^2} &\approx&
 \frac{1}{n_s^2+\nu|\phi_s|^2}\left. \frac{\partial r}{\partial x}\right|_{\epsilon^-}
  -\frac{1}{n_l^2}\left. \frac{\partial r}{\partial x}\right|_{\epsilon^+}\nonumber\\
 &+&\int_{-\epsilon}^{+\epsilon}\!\!dx\frac{\partial_x^2r}{n_0^2+\nu|\phi|^2} = 0
 \label{SIM4}
\eea
 in the limit of $\epsilon \rightarrow 0$. These considerations thus reduce \E{SIM3.5} to
\be
\Delta r \equiv r_s-r_L=2C_s \text{ln}\left(\frac{n_s^2+\nu|\phi_s|^2}{n_l^2}\right)
\label{latent_heat_analytic}
\ee
\EE{latent_heat_analytic} predicts a temperature rise as an undercooled liquid orders its density across the solid-liquid interface. This prediction can directly be compared to the results in Fig.~\ref{fig:slab_thermal}, where we observe $\Delta r_{\rm simulation}=0.00166$. For the given parameters, the 1-mode approximation calculation from the phase diagram calculation predicts $\phi_s=0.09466$. Substituting this in \E{latent_heat_analytic} yields $\Delta r_{simulation}=0.00159$, which is $4\%$ deviation from the simulation result, in excellent agreement with the analytical derivation of \E{latent_heat_analytic}.

The PFC temperature change $\Delta r$ can be more closely related to the latent heat source term in the {\it Model C} type of phase model for thermally controlled solidification \cite{Hoh77}. In standard Model C phase field model, the starting point is the heat equation of the form
\be
\dot{T}=\alpha\nabla^2T+L\frac{h'(\phi)}{C_p}\frac{\partial\phi}{\partial t}
\ee
We rescale this equation to PFC units with $T(r)=T_0\theta(r)$, yielding
\be
\dot{r}=\alpha\nabla^2r+\alpha\frac{\nabla \theta'\cdot\nabla r}{\theta'}+\frac{L}{T_0\theta'}\frac{h'(\phi)}{C_p}\frac{\partial\phi}{\partial t}
\ee
Neglecting the cross terms and going through a similar procedure as above gives,
\be
\Delta r=\frac{L}{T_0 \theta' C_p}
\ee
Identifying this expression with the Latent heat expression in PFC model gives
\be
L/(T_0C_p)=\Delta r\theta' =2C_s  \theta' \text{ln}\left(\frac{n_s^2+\nu\phi_s|^2}{n_l^2}\right)
\ee
Note that mapping coefficients aside, this expression depends on both the average density and the amplitude. In the traditional phase field limit, there is no density change (i.e. $n_s=n_l$), and so
\be
L/(T_0C_p)=2C_s  \theta' \text{ln}\left(1+\frac{\nu|\phi_s|^2}{n_l^2}\right)\approx 2 \nu C_s  \theta'  \, \frac{|\phi_s|^2}{n_l^2}
\label{latent_heat}
\ee
Equation~(\ref{latent_heat}) makes manifest that Latent heat release is proportional to density change due to ordering of the liquid into solid.

The approach described in this subsection shows the correspondence of our formalism with the model C type dynamics on long length scales, while incorporating  new features in thermal transport that arise solely from the properties of the PFC model at the atomic scale (e.g. density re-arrangements, defect flows, and other atomic-scale features attainable by the PFC density field). Experimentally relevant thermal traces of microstructure rearrangements, like recalescence for example, can therefore be observed using this formalism. 

To directly apply our modelling formalism to experiments, the basis of temperature itself still requires more investigation. In the basic PFC model, temperature is buried implicitly in the $r$ parameter. It is not clear how the $\theta$ function should relate it to the thermodynamic temperature $T$. Moreover, even with an appropriate $\theta$ function, due to the approximated form of the PFC free energy it isn't clear if the model can reproduce the right physics across the whole  phase space from vapour to solid. One solution is to restrict the investigation to a small part of phase space as has been done in the past. In what follows,  we propose a new and quite different approach aimed at improving the thermodynamic consistency of  PFC modelling, and that covers a wide range of a material's phase space.

\section{Towards a unified structural PFC model with a physical temperature scale}
\label{newmodel}
Phase field crystal type models introduced to date have all contributed to making PFC models and their dynamics increasingly more consistent with thermodynamics. Two key features are, however, still lacking. The first  deals with relating phenomenological PFC model parameters to  thermodynamic temperature. For example, in the last section, we assumed that  the parameter $r$ in the original PFC model can be mapped to thermodynamic temperature through an unknown function $\theta(r)$, i.e., $T=T_0\theta(r)$. Several PFC models  have presumed such relationships to interpret  results in localized regions of model-specific phase diagrams with measurable quantities \cite{elder2004,jaatinen2009,karma2013,Asadi2015}. In this section, we go beyond previous approaches, and derive a general PFC-type formalism that explicitly relates model parameters to thermodynamic temperature, and which integrates naturally with the thermo-density formalism introduced in the previous sections. 
The second feature lacking in PFC modelling is a  form for the free energy that is  both tractable in its implementation and quantitative enough to incorporate phase changes over a robust range of density temperature, and pressure space.  Moreover, it is crucial to address this issue while   allowing for  the stabilization of complex crystalline structures, and for {\em efficient dynamical simulations} of phase transformations involving crystalline and disordered phases.  

In the remainder of this section, we propose a new density functional formalism that addresses the above problems, leading to a PFC-style model that takes a significant step closer to a unified and quantitative PFC theory for pure materials. .

\subsection{A New Density Functional Approach: Expanding around the Van der Waals fluid}
One of the key simplifications in PFC models is the expansion of the ideal [non-interacting] free energy. This approach makes the model analytically tractable and efficient to simulate. However, it severely limits the shape of the phase diagram, and also precludes a full solid-liquid-vapor description of a material's phase space. This is particularly felt near zero density, where the influence of the logarithm terms can lead to very low density phases (e.g. voids). 

Here, we propose a model that starts with the full Van der Waals free energy to model long range (mean field) interactions in a system, complemented with multi-point  correlations  designed to capture short-range interactions and the emergence  of solid phases. 
In particular, we break up the excess interaction energy of standard CDFT as a sum of a Van der Waals term and other excess effects, namely,
\begin{equation}
{F}[\rho,T]={F}_{id}[\rho,T]+{\Phi_{VdW}}[\rho,T]+{\Phi}[\rho,T],
\label{free_general2}
\end{equation} 
where 
\begin{equation}
{F}_{id}[\rho,T]=\int d \bx \, \rho k_BT\left(\ln\left(\lambda^3\rho\right)-1\right)
\end{equation}
and
\begin{equation}
\!\!{\Phi_{VdW}}[\rho,T]\!=\!- \!\int \!d \bx  \, k_BT\left\{\rho \ln\left(1 \!- \! \tilde{b}\rho\right)+\frac{\tilde{a}}{k_B T}\rho^2\right\}
\end{equation}
The Van der Waals theory works well as its mean field limit is easy to analyze and can accurately describe disordered phases ~\cite{plischkebergersen}. Note that we have named the Van der Waals parameters $\tilde{a}$ and $\tilde{b}$ in anticipation that these will be rescaled and renamed to $a$ and $b$. The excess term $\Phi[\rho,T]$ is responsible for all short range interactions not captured by $F_{id}[\rho,T]+F_{VdW}[\rho,T]$.  By performing a functional expansion around a uniform reference density $\bar{\rho}$ corresponding to an equilibrium fluid phase, we arrive at the following the free energy,
\bea
F&=&\bar{F}[\bar{\rho},T]+\Phi\left[\bar{\rho},T\right] + \int d\bx \, (\bar{\mu}-V_{ext}(\bx))\delta{\rho}(\bx)\nonumber\\
&+&\int d\bx k_BT\left\{ \rho \ln\left(\frac{\rho}{\bar{\rho}}\right)-\delta\rho+\rho \ln\left(\frac{1-\bar{\rho}\tilde{b}}{1-\rho\tilde{b}}\right)\right.\nonumber\\
&-&\left.\delta \rho\left(\frac{\bar{\rho}\tilde{b}}{1-\bar{\rho}\tilde{b}}\right)-\frac{\tilde{a}}{k_BT}\delta\rho^2 \right\} \label{firstVdW} \\
&-&\frac 1 2 \iint d\bx \, d\bx' \, k_BT \, \delta \rho(\bx)\tilde{C}^{(2)}(\bx,\bx',T)\delta \rho(\bx')+ \cdots, \nonumber
\eea
where $\delta \rho=\rho-\bar{\rho}$, $V_{ext}$ is an external field, $\bar{\mu}$ is the chemical potential of the reference state, and 
\bea
\bar{F}[\bar{\rho},T]&=&\int d \bx  k_BT\left\{ \bar{\rho} \ln\left(\lambda^3\bar{\rho}\right)- \bar{\rho}-\bar{\rho} \ln\left(1-\tilde{b}\bar{\rho}\right) \right. \nonumber\\
&& \left. +\frac{\tilde{a}}{k_B T}\bar{\rho}^2\right\}
\eea
is the free energy at the reference density. At this stage, we have identified by $\tilde{C}^{(2)}(\bx,\bx',T)$ as the general form of the direct two-point correlation function generated by the excess energy. We have denoted by three dots $[\cdots]$ the presence of higher order correlations which we have not written down as they are similarly generated by direct correlation functions. These will be assumed to act only on short wavelengths.

To proceed further, we introduce the reduced density, $n=(\rho-\bar{\rho})/\bar{\rho}$ or $\rho=\bar{\rho}(n+1)$, and the length scaling $\br=\bx/R$, where R is a characteristic length scale, typically a lattice constant of one of the solid phases. In what follows, we will only consider the free energy difference relative to the reference state. Thus, we will work with $\Delta F=F-\bar{F}[\bar{\rho},T]-\Phi\left[\bar{\rho},T\right]$.  For completeness, we keep the terms linear in density, and define a rescaled chemical potential $\tilde{\mu}=\bar{\mu}/K_BT_0$ and  external potential $\tilde{V}_{ext}=V_{ext}/k_BT_0$. (While linear terms have an important role to play in the system pressure~\cite{ZhifengNewPaper}, they do not change the essence of the phase diagram.)  Here, $T_o$ is some reference temperature. Combining the above elements gives a complete form of the proposed  density functional,
\bea
\mathcal{F}&=&\int d\br \left[\tilde{\mu}-\tilde{V}_{ext}-\left(\frac{\tau}{1-b}\right)\right]n\nonumber\\
&+&\int d\br \, \tau\left\{ \left(n+1\right) \ln\left(\frac{\left(n+1\right)(1-b)}{1-\left(n+1\right)b}\right)-\frac{a}{\tau}n^2 \right\} \nonumber\\
&-& \frac 1 2 \iint d\br \, d\br' \tau \,n(\br)C^{(2)}(\br,\br',\tau)n(\br') + \cdots,
\label{cdftmodel}
\eea
where we have defined $\mathcal{F}={\Delta F}/{ k_BT_0 \bar{\rho}R^d}$, $b=\bar{\rho}\tilde{b}$, $a=\bar{\rho}^2\tilde{a}/k_B T_0$, the rescaled temperature $\tau\!=\!T/T_0$ and $C^{(2)}= \bar{\rho}  R^d \tilde{C}^{(2)}$. The first and second lines in \E{cdftmodel} contain effects from the ideal gas and the Van der Waals interactions, the third line contains all the supplementary correlations required to  stabilize solid phases. 

\EE{cdftmodel} is still incomplete as we are missing a form for the correlation function $C^{(2)}$ (and higher order terms in the $[\cdots]$) to create a full phase diagram. In practice, we want to affect short range ordering (e.g. solid phases) without affecting the long-wavelength behaviour of the first two lines of~(\ref{cdftmodel}). To proceed, we  therefore separate the length scales of the Van der Walls terms.  Effectively, this amounts to  enforcing the Van der Walls free energy terms only at long wavelengths by making them explicit functions of $n_{mf}(\bf r) \! \equiv \! \int\text{d}\br\chi(\br-\br')n({}\bf r')$, instead of the microscopically varying $n(\bf r)$ field.  This gives,
\bea
\mathcal{F}&=&\int d\br \left[\tilde{\mu}-\tilde{V}_{ext}-\left(\frac{\tau}{1-b}\right)\right] n_{mf}\nonumber\\
&+&\int d\br \, \tau  \left\{ \left(n_{mf}+1\right) \ln\left(\frac{\left(n_{mf}+1\right)(1-b)}{1-\left(n_{mf}+1\right)b}\right)-\frac{a}{\tau}n_{mf}^2 \right\} \nonumber\\
&-&\frac 1 2 \iint d\br \, d\br' \, \tau \,n(\br)C^{(2)}(\br,\br',\tau)n(\br')+\cdots
\label{cdftmodel2}
\eea
Since the first line in~(\ref{cdftmodel2})   essentially  maintains its mean field form, it will ensure that the long wavelength behavior of the model follows the correct Van der Walls thermodynamics. In proceeding as above, we have tacitly assumed that the short wavelength interactions removed from the original Van der Walls terms are subsumed into $C^{(2)}$ and the higher order correlation terms denoted by $[\cdots]$.  This approach also avoids the  severe constraint of requiring a positive density  at the atomic scale. which forces the solid peaks to become very sharp in full CDFT calculations, and effectively kills the ability for {\it dynamical microstructure simulations, the quintessential advantage of PFC modelling}. In other words,  the long wavelength behaviour of the density in \E{cdftmodel2} will  follow the correct Van der Walls thermodynamics, but the model still allows for smooth atomic-scale density oscillations that go both above and below the reference density, thus making it possible to model crystalline patterns in a computational and analytically tractable manner.  

We proceed next to simplify $C^{(2)}$ by breaking it into a structural term $C^{(2)}_{struct}$ that is used to stabilize crystalline states, and the other short-range correlations that now become subsumed into the $[\cdots]$ terms. Here, we employ an XPFC type kernel \cite{greenwood2010}. Namely, we define $C^{(2)}_{struct}$ as
\be
\!\!C^{(2)}_{struct}(\bk)\!=\!B_xe^{-T/T_0}\hat{C}_\bk\!=\!B_xe^{-T/T_0} e^{-(\bk-\bk_0)^2/2\alpha^2}
\label{xpfc_c2}
\ee
where $B_x$ is a constant, $k$ is the magnitude of $\bk$ and $k_o$ is the magnitude of $\bk_0$, which  defines the equilibrium reciprocal lattice vector of the first Bragg reflection of an 2D HCP lattice or a 3D BCC lattice. The single peak XPFC formalism has the advantage of allowing for independent tuning of the lattice constant, elastic modulus or surface energetics due to the rapid decay of the correlation peak. Adding multiple Gaussian  type peaks such as those in \E{xpfc_c2} to $C^{(2)}_{struct}$ allows the model to describe  different structures. Formally, there is one peak representing  the main Bragg reflection peak from a set of crystal  planes; in practice, only the first few [smallest] $k$-peaks are retained  (e.g., one for 2D HCP and 3D BCC, two for 2D squares and 3D FCC, three for 3D HPC 3D \cite{GREENWOOD11}, etc). We can also expand to 3-point correlations following the recent work of Seymour et al~\cite{Matt,Matt_thesis,Alster2017B} in order to extend the range of crystalline structures to non-metallic materials. It is noted that the use of higher order correlations typically requires that we go beyond 1-mode analysis of the free energy in order to retain the accuracy of simpler 1-mode approximations used for simpler structures. In this work, we will demonstrate the new model and its thermo-density coupling using a 1-peak $C^{(2)}_{stuct}$ function.
 
The original XPFC models parametrized temperature by $e^{-\sigma^2/\sigma_0^2}$, where $\sigma$ is used as a model temperature parameter. Here, we follow the approach of Alster et. al. \cite{ALSTER17} who effectively define $\sigma^2\sim T/T_o$. This is choice is more consistent with the temperature dependence of the Debye-Waller factor observed in experiments. The reference temperature $T_0$ here can in principle be   different from that used to scale the density functional expansion above. In this work, we aim to show that the model proposed in this section is physically consistent {\it and } robust enough to cover a wide range of material systems quantitatively. Thus, for convenience, we use the same reference temperature $T_0$ here as previously introduced to rescale the free energy functional. 

The model of \Ess{cdftmodel2}{xpfc_c2} is still missing the higher order correlation terms separated out of the Van der Walls terms and buried in the $[\cdots]$ terms. In terms of a density expansion of the form $n=n_0+\phi\sum_\textbf{G} e^{i\textbf{G}\cdot\br}$ the first two lines in~(\ref{cdftmodel2}) only give, to lowest order, average density  ($n_0$) contributions. On the other hand, substituting \E{xpfc_c2} in the third line of~(\ref{cdftmodel2}) also  gives a $\phi^2$ amplitude contribution to the mean field  of the theory, which only contributes to the free energy of the solid phase where $\phi>0$. However, we still require a~$\phi^4$ theory to stabilize solid phases, and coupling terms between $n_0$ and $\phi$ to control the average density of the solidus. Another way to look at this is that we have enforced Van der Waals Theory (a complete theory), and crystallographic ordering at short wavelengths, but important thermodynamics at short wavelengths is still missing. 

\begin{figure}
\centering 
\includegraphics[width=0.45\textwidth]{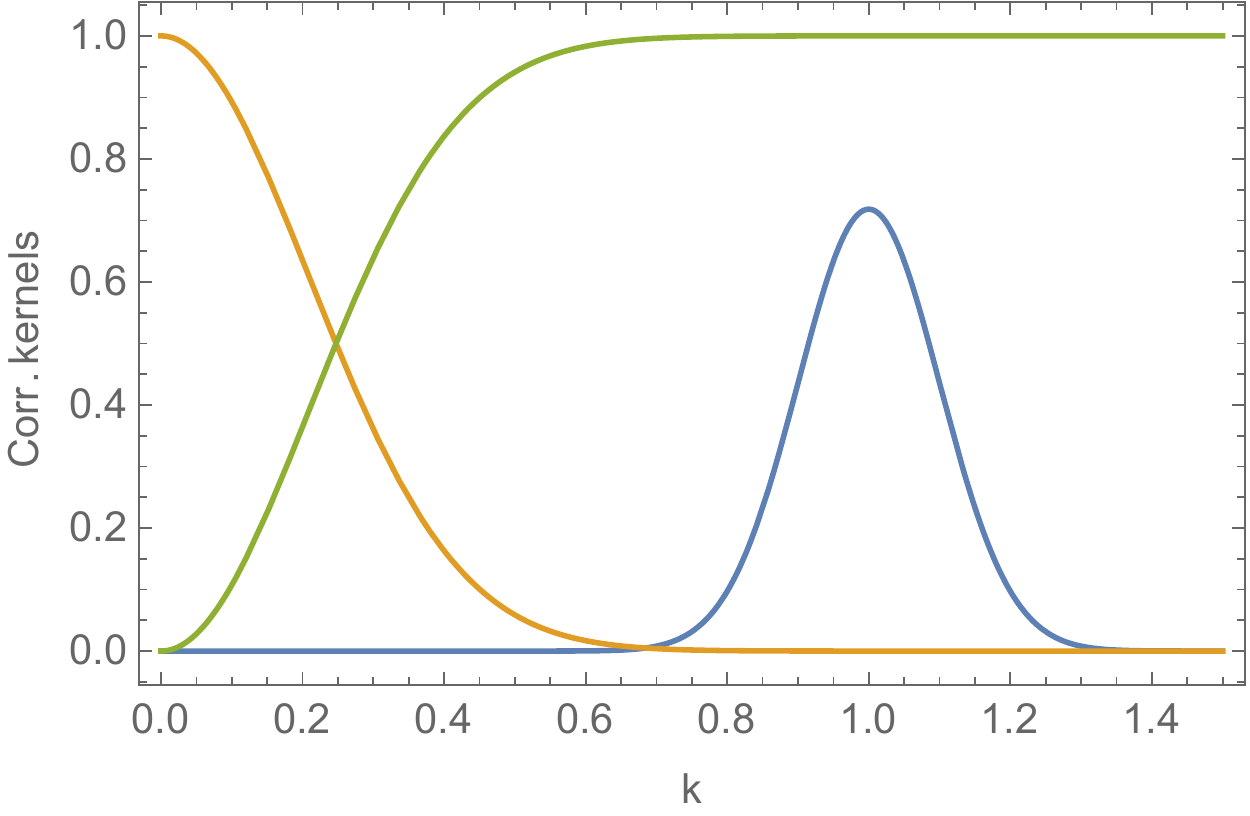}
\caption[Plot of the 2-point correlation kernels at play in the model, in reciprocal space.]{Plot of the 2-point correlation kernels  introduced in the model defined in \E{cdftmodel3}, in reciprocal space. Blue: XPFC kernel~(\ref{xpfc_c2}), Orange: $\chi^{(2)}(\bk)$ smoothing kernel used to create the $n_{mf}$ appearing in~(\ref{cdftmodel3}), Green: $\zeta^{(2)}(\bk)$ kernel modulating the short range correlation functions in \E{cdftmodel3}). Parameter values used in plot: $B_x=3.5$, $T/T_0=2.5$, $\alpha=0.1$, $k_0=1$, $\lambda=1$.\label{fig:kernels_unified}}
\end{figure}

To remedy this, we  introduce a series of multi-point correlation functions that only affect the short wavelength behaviour. These emerge from the $[\cdots]$ terms in \E{cdftmodel2}. For this, we propose a set of correlations denoted $\zeta^{(m)}$, which are defined as products of two-point functions,  given by
\be
\zeta^{(m)}\!=\!\delta(\br_1-\br_2) ... \delta(\br_1-\br_{m})-{\chi}(\br_1-\br_2)...{\chi}(\br_1-\br_{m}),
\label{zeta_defII}
\ee
 where  ${\chi}(k)\!\!=\!\!\text{exp}(-k^2/(2\lambda^2))$ are long-wavelength kernels, defined in reciprocal space. (For simplicity, in this work we take the $\chi(\br_1 \! - \! \br_m)$ functions in \E{zeta_defII} to be the same as the $\chi$ functions defined previously). In terms of the $\zeta^{(m)}$  functions, the final form fo the model is written as
\bea
\mathcal{F}&=&\int d\br \left[ \tilde{\mu}-\tilde{V}_{ext}- \left(\frac{\tau}{1-b}\right)\right] n_{mf}\nonumber\\
&+&\int d\br \, \tau\left\{ \left(n_{mf}+1\right) \ln\left(\frac{\left(n_{mf}+1\right)(1-b)}{1-\left(n_{mf}+1\right)b}\right)-\frac{a}{\tau}n_{mf}^2 \right\} \nonumber\\
&-&\frac 1 2 B_x \! \int d\br\,  \tau(\br) e^{-\tau} n(\br) \mathcal{F}^{-1} \!\! \left\{ \int d\bk'  \hat{C}_{\bk^\prime} n(\bk') \right\} \label{cdftmodel3} \\
&+&\! \! \sum_{m=1}^{4} \frac{a_m}{m}  \left(\int d \br_1..d\br_m\zeta^{(m)}(\br_1,..,\br_m)n(\br_1)..n(\br_m)\right) \nonumber
\eea
The last line in~(\ref{cdftmodel3}) comprises a series of correlation functions defined on short wavelengths, each weighted by dimensionless $\tau$-dependent coefficients $a_m$. As an example, $\chi^{(2)}(\bk)$ contributes only at long wavelengths, thus  acting as a low pass filter, but  $\zeta^{(2)}(\bk)=1-\chi^{(2)}(\bk)$ contributes at short wavelengths, thus essentially acting as high pass filter.  The $\zeta^{(m)}$ correlation terms allow for control over the solid phase density-temperature properties in the phase diagram. The width of these short-range correlation functions also impacts the interface energy and solid compressibility. The $\hat{C}_{\bk}(\bk)$ correlation function selects the crystal structure and also tunes the solid interfacial energy or elastic modulus. Long wavelength correlations are controlled by the Van der Walls theory. 
Figure \ref{fig:kernels_unified} illustrates the shape of all 2-point kernels appearing in \E{cdftmodel3}, in reciprocal space.  
The study of interface energy with the model in \E{cdftmodel3} will be the topic of future work.


\subsection{Equilibrium properties  of the model }
\begin{figure}[h!]
\centering
\includegraphics[width=0.45\textwidth]{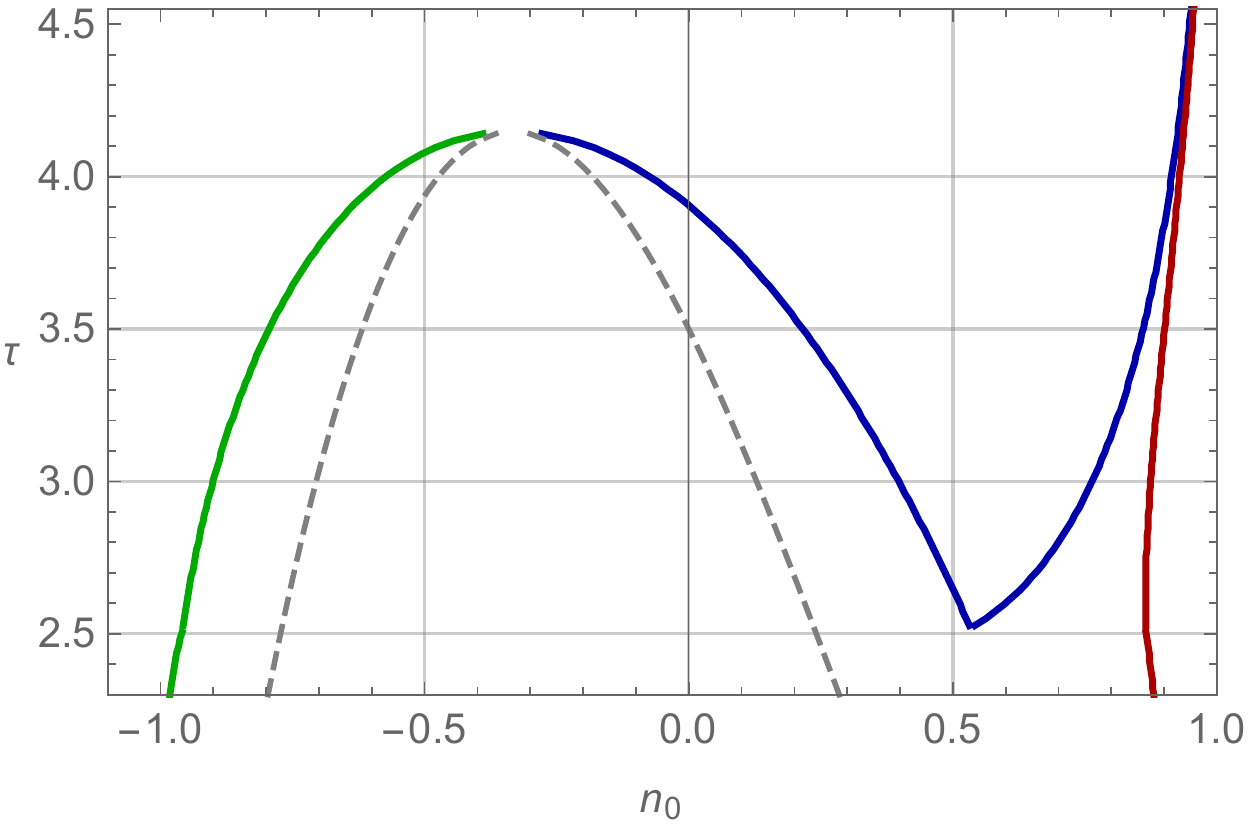}
\caption[Phase diagram of the full model.]{Typical phase diagram of the model in \E{cdftmodel3}. Green, blue and red are, respectively, the coexistence densities of vapor, liquid and periodic phases. Parameter values: $a=7$, $b=0.5$, $B_x=3.5$, $a_2=5.45$, $a_3=-2$, $a_4=0.1$. \label{fig:PD1}}
\end{figure}

Mean field calculations of the model are done by using a 1-mode approximation approach. The long-wavelength correlation kernels $\chi$ and $\zeta$ are approximated as delta functions in $\bk$ space. The phase diagram is computed by the usual common tangent or Maxwell equal area construction. To focus on demonstrating the salient features of the model in this work, we arbitrarily set $\tilde{\mu}=\tau/(1-b)$ and $V_{ext}=0$, effectively cancelling out all the linear terms. A resulting phase diagram is shown in Fig.~\ref{fig:PD1}. It features both a critical and a triple point, and a strongly asymmetric vapor/liquid phase separation as one expects from the Van der Waals theory. In addition, there are both liquid-solid and vapor-solid coexistence regions, which can be made quantitive by controlling the $a$, $b$ and $T_0$ parameters. It is also noteworthy that the density axis for the uniform phases cover a physical range, starting at $n_o = -1$ (i.e. $\rho=0$). The phase diagram also contains a solid phase, which has, for this choice of model parameters, a density $\sim 25\%$ higher than the liquid density near the triple point and approaches the liquid density at high temperature. To our knowledge this is the first PFC model to describing solid-liquid-vapour phases over a broad, and physically consistent,  ranges in both temperature and density.

\begin{figure}
\centering
\includegraphics[width=0.45\textwidth]{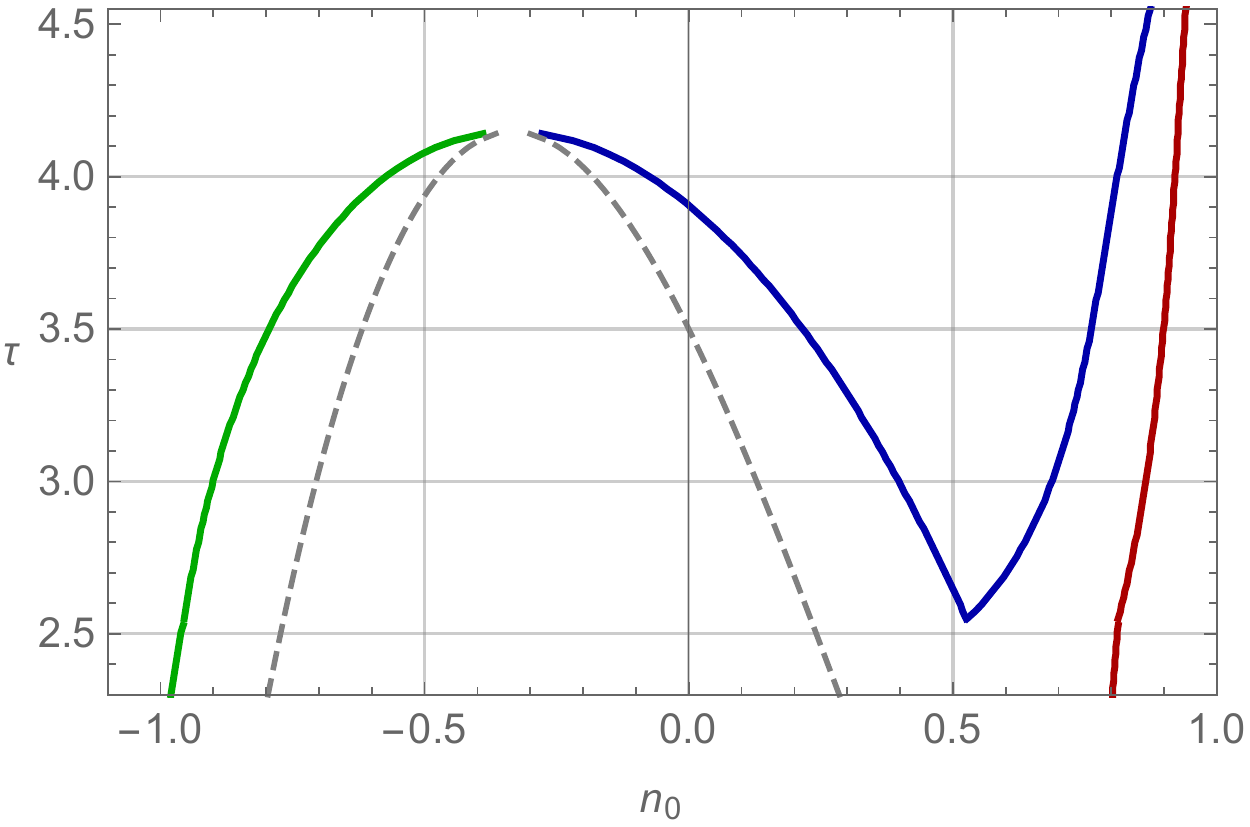}
\caption[Phase diagram of full model, with temperature dependent $a_2$ and $a_3$ coefficients in the non-structural correlation terms.]{Phase diagram of the the model in \E{cdftmodel3}, with temperature dependent $a_2$ and $a_3$ coefficients in the non-structural $\zeta^{(m)}$ correlation terms. Green, blue and red are, respectively, the coexistence densities of vapor, liquid and periodic phases. The dashed lines show the instability boundary of phases. Same parameter values as fig.~\ref{fig:PD1}, except $a_2=-2.26082 + 2.41018 \tau + 0.06 \tau^2$ and $a_3=0.689182 - 1.08982 \tau + 0.06 \tau^2$. \label{fig:PD2}}
\end{figure}

It is seen in Fig.~\ref{fig:PD1} that using constant $a_m$ factors reduces the solid-liquid density jump rapidly as temperature is increased. It possible to obtain a larger solid-liquid density jump, and to decrease the divergence of the solid-liquid phase boundaries by introducing a temperature dependence in the $a_m$ coefficients.  In this study, we used a second order polynomial in $\tau$ for $a_2$ and $a_3$. We determine the parameters of the $\tau$-expansion of $a_2$ and $a_3$ by interpolating between their desired values at the lowest considered temperature $\tau_l$ (i,e. $a_{2,l}$, $a_{3,l}$) and highest considered temperature $\tau_h$ (i.e. $a_{2,h}$, $a_{3,h}$), while the quadratic term pre-factor $Q$ is fixed empirically for demonstration purposes. The specific form of the expansion for $a_2$ and $a_3$ are given by,
\bea
\!\!\!a_2(\tau)=Q \tau^2 &-& \frac{ (-a_{2,h} \!+\! a_{2,l} \!+\! Q \tau_h^2 \!-\! Q \tau_l^2)}{\tau_h \!-\! \tau_l}\, \tau \nonumber\\
&-& \frac{-a_{2,l} \tau_h + a_{2,h} \tau_l \! -\! Q \tau_h^2 \tau_l \!+\! Q \tau_h \tau_l^2}{\tau_h \!-\! \tau_l} \nonumber \\
\!\!\!a_3(\tau)=Q \tau^2 &-& \frac{ (-a_{3,h} \!+\! a_{3,l} \!+\! Q \tau_h^2 \!-\! Q \tau_l^2)}{\tau_h \!-\! \tau_l}\, \tau \label{temp_fits}\\
&-& \frac{-a_{3,l} \tau_h + a_{3,h} \tau_l  - Q \tau_h^2 \tau_l + Q \tau_h \tau_l^2}{\tau_h \!-\! \tau_l} \nonumber
\eea
A phase diagram with these $\tau$-expanded correlation coefficients in the new PFC model is shown in figure~\ref{fig:PD2}, where the addition of this dependence has allowed us to widen the solid-liquid  coexistence region.   
More careful fitting of $\tau_l$ and $\tau_h$ and their corresponding $a_{2,l}$, $a_{3,l}$, $a_{2,h}$, $a_{3,h}$  can  expand the density jump further, but the form of the short-range correlations introduced cannot lead to completely parallel solid-liquid density lines, although not all materials have parallel solid-liquid coexistent lines.

The   limitation in further expanding the solid-liquid density jump at high temperature arises due to the incompressibility of the liquid inherent in the Van der Waals formalism. In particular, the logarithm term in~(\ref{cdftmodel3}) diverges as the average density approaches $n^*=(1-b)/b$ (corresponds to $n_0=1.5$ in the figures shown). This incompressibility limit is due to the excluded volume introduced in the Van der Waals theory, and is beneficial for modelling the uniform phases in the PFC model, but does not accommodate phases at arbitrarily higher density phases, namely the periodic (solid) phases.  

\subsection{Expansion of the Van der Walls terms}

To model a larger range of solid-liquid density jumps in the proposed model, one can either relax the exclusion volume at higher density, or simply expand the uniform phase free energy. We proceed by expanding the logarithmic terms in \E{cdftmodel3}, to recover a more flexible polynomial expansion  of the free energy functional. The drawback of this approach is that the  sharp cusp of the free energy at low density becomes less accurate. Higher order expansions or more elaborate fitting techniques such as a spline fitting could be applied to both accurately fit the sharp cusp at low density while simultaneously allowing a smooth description of liquid at high densities. These fitting techniques will not be studied here. Our aim here is to show how to create a complimentary version of the unified model that offers flexible control of the high density phases and their properties; the unified model of \E{cdftmodel3} can be used where the accuracy of the vapour phases is required.

\begin{figure}[h!]
\centering
\includegraphics[width=0.45\textwidth]{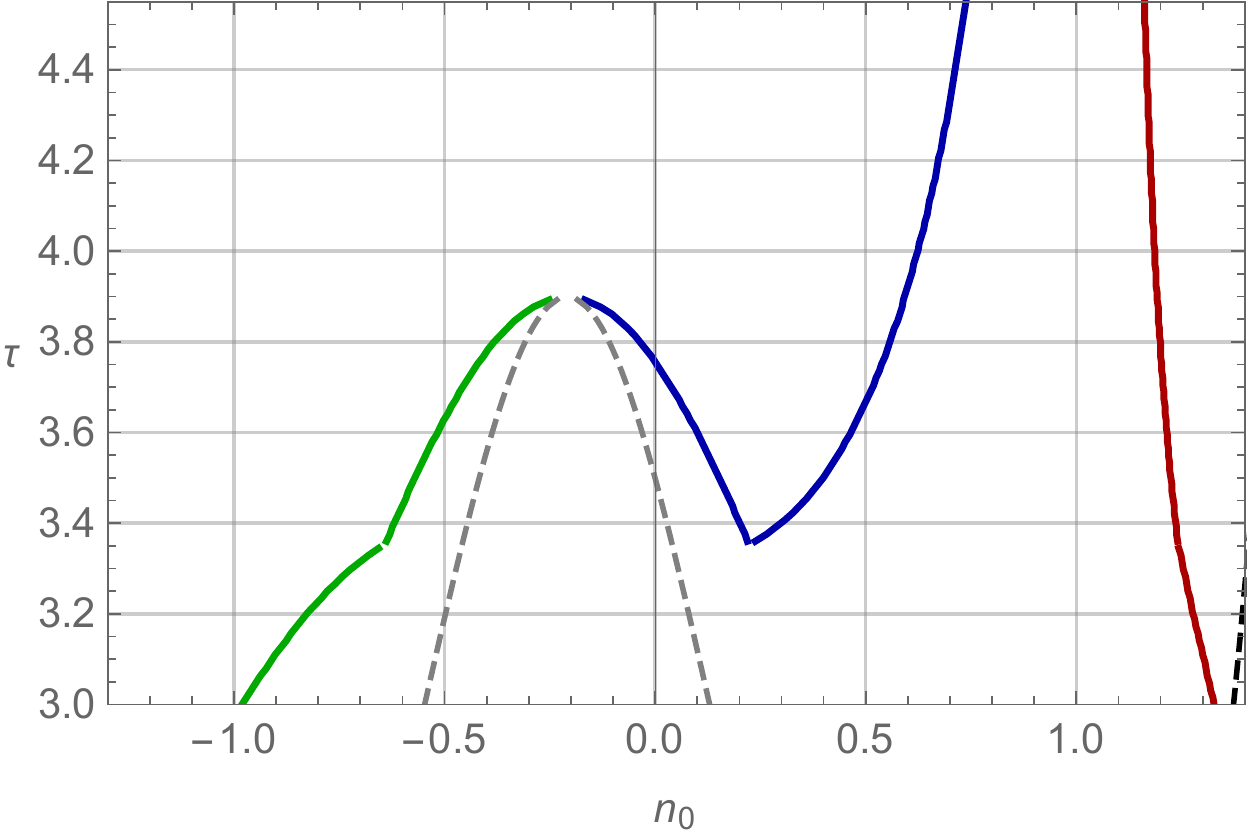}
\caption[Phase diagram of the expanded 4th order model.]{Phase diagram of \E{cdftmodel3} with logarithm  terms expanded to 4th order. Green, blue and red are, respectively,  the coexistence densities of vapor, liquid and periodic phases. The dashed lines show the instability boundary of phases. Same parameter values as fig.~\ref{fig:PD1}. \label{fig:PD4}}
\end{figure}

We start by Taylor expanding the uniform part of the free energy in \E{cdftmodel3} to fourth order. The expansion is taken around $n_0=0.05$ to match the true free energy versus $n_o$ of the full model as closely as possible across the temperature range we are interested in. Figure~\ref{fig:PD4} shows the corresponding phase diagram. It features a  wider liquid/solid density jump, but a significantly less asymmetric liquid and vapor coexistence. The parameters used are the same as those in Fig.~\ref{fig:PD1}. 

\begin{figure}[h!]
\centering
\includegraphics[width=0.45\textwidth]{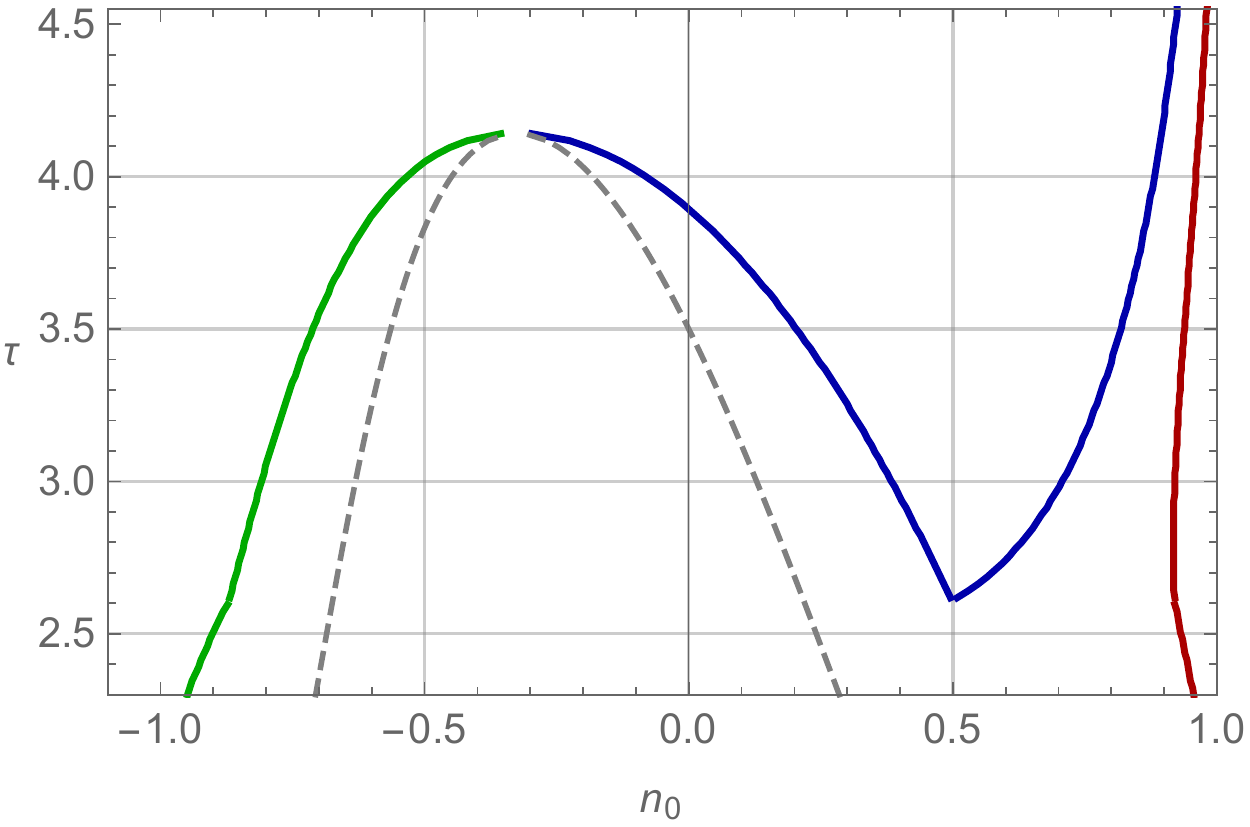}
\caption[Phase diagram similar to Fig.~\ref{fig:PD2F_10} with a 10th order expansion of the logarithm terms in \E{cdftmodel3}. ]{Phase diagram corresponding to that in Fig.~\ref{fig:PD1} with a $10^{th}$ order expansion of the uniform free energy logarithm terms in \E{cdftmodel3}. Notice the vapor-liquid phase separation starts to take a more asymmetric shape. \label{fig:PD2_10}}
\end{figure}

\begin{figure}[h!]
\centering 
\includegraphics[width=0.45\textwidth]{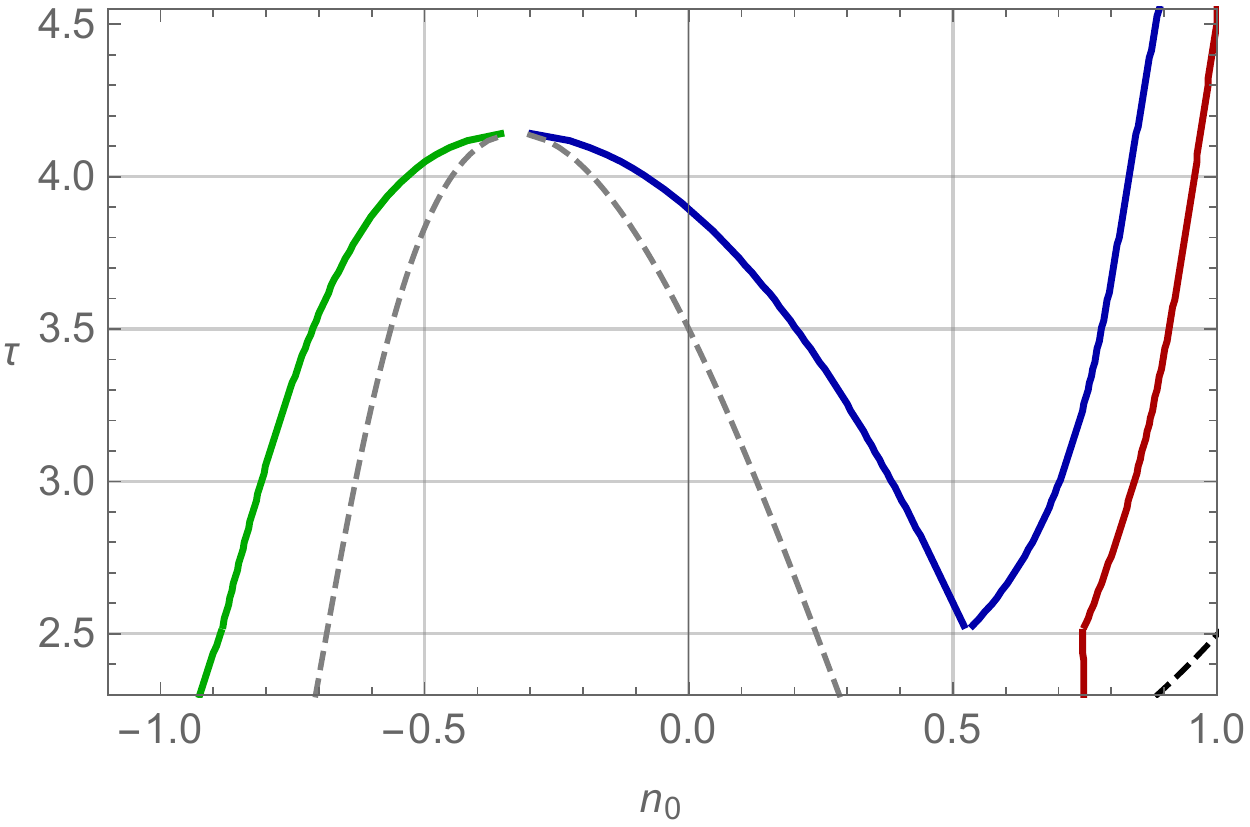}
\caption[Same system as in  fig.~\ref{fig:PD2_10} but with a temperature dependance in the $a_2$ and $a_3$ coefficients.]{Same system as in fig.~\ref{fig:PD2_10} but with a temperature dependance in the $a_2$ and $a_3$ coefficients. The density jump and slope of the liquid/solid coexistence lines are now more parallel and habe smaller slopes. Here, $a_2=-1.3825 + 1.66 \tau + 0.05 \tau^2$ and $a_3=0.572045 - 0.885455 \tau + 0.05 \tau^2$. 
\label{fig:PD2_10_b}}
\end{figure}

To  improve on this behaviour, we perform a 10th order Taylor expansion of the uniform free energy around $n_0=0.05$. Both Vapor and Liquid free energy wells are now more accurately captured, and  leads to a low-density part of the phase diagram very close to the one in Fig.~\ref{fig:PD1}, as shown in Fig.~\ref{fig:PD2_10}, although some discrepancy is still seen at lower temperatures due to the expanded nature of the polynomial.   Note also that the solid-liquid density jump now becomes wider and more parallel at higher temperatures. 
To further expand the solid-liquid density jump, we can again assume a temperature dependance of $a_2$ and $a_3$ given by \Es{temp_fits}. By considering the effect of these parameters  on the free energy, we can obtain a wider and more parallel liquid and solidus coexistent density lines. A typical example is shown in Fig.~\ref{fig:PD2_10_b}. Note that the slope of the liquid/solid coexistence line is smaller than the previous phase diagrams; this is done to demonstrate the model's flexibility.

\begin{figure}[h!]
 \centering
 \includegraphics[width=0.45\textwidth]{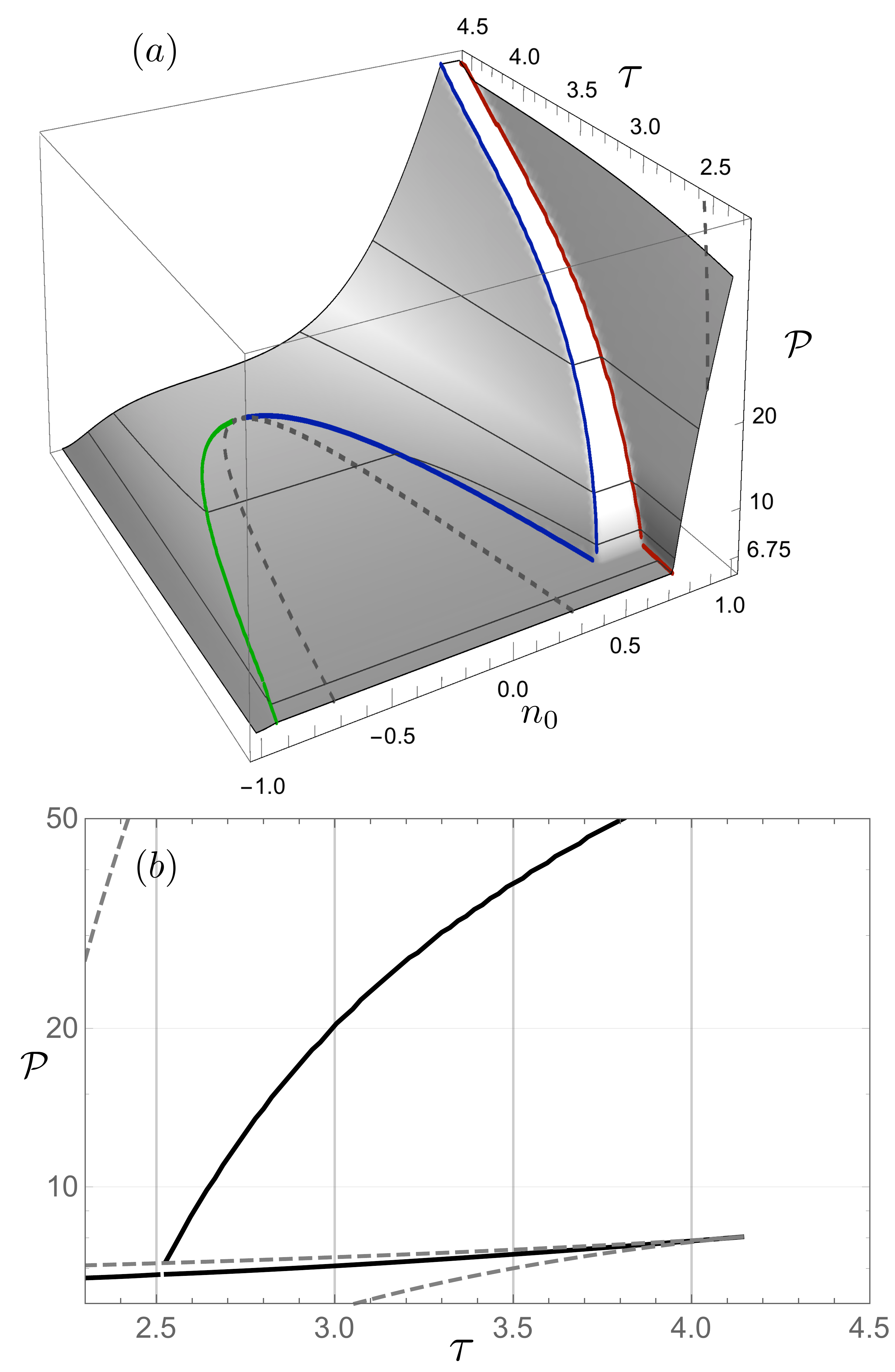}
 \caption[Phase diagram of the full model.]{(a) Same phase diagram as in Fig.~\ref{fig:PD2_10_b}, showing the reduced pressure-temperature view.   The dashed lines show the instability boundary of phases. (b) Pressure-Temperature phase diagram of the system in Fig.~\ref{fig:PD2_10_b}. It features  vapor, liquid and solid phases as well as coexistence lines. Dashed lines are metastability lines.} \label{fig_pressure}
 \end{figure}

Fig.~\ref{fig_pressure}(a) shows the density-temperature-pressure phase diagram corresponding to the system in Fig.~\ref{fig:PD2_10_b}. The units of $[\mathcal{P}]=P/k_B \bar{\rho}T_oR^d$. The figure features solid-liquid, solid-vapor and vapor-liquid coexistence regions, and is in excellent qualitative agreement with experimental phase diagrams for pure materials. The Pressure-Temperature phase diagram (Fig.~\ref{fig_pressure}(b)) also shows a behaviour consistent with experiments. Along with the equilibrium phase boundaries, Fig.~\ref{fig_pressure}(b) also shows analytical estimates for the metastability regions of the different phases (dashed lines). Transforming from a metastable to stable phase requires a nucleation event. Crossing the metastable boundaries is associated with the appearance of an unstable wavelength. 

In general, once the parameters setting the characteristics of the uniform phases are determined, other parameters can then independently sought to set the elastic modulus (solid), surface energies, compressibility, etc.  Using a 4th order expansion provides the barebones structure for a somewhat qualitatively correct phase diagram. The 10th order expansion provides a much more accurate description of the entire material system. Higher order expansions or different fitting techniques can be used to match the vapor/liquid phase diagrams in the regions of interest. Moreover, the use of the $\zeta^{(m)}$ correlation functions gives a new degree of control of the liquid/solid {\it and} solid state thermodynamics, without affecting the the selection of crystallographic symmetries or the vapor/liquid properties. Furthermore, making the coefficients of the $\zeta^{(m)}$ functions have temperature dependence can provide added flexibility to match both the slope and width of the liquid/solid coexistence in a wide range of materials.

The  phase field crystal type of theory derived in this section, i.e., \E{cdftmodel3} and its variants that employ an expanded form of the uniform phase free energy,  lead to an accurate description of  3 phase equilibrium in materials whose triple and critical points aren't too far from one another in temperature. The proposed formalism makes it  possible to model both critical point phenomena and very low density phases, solidification of crystalline phases from liquid or vapour, as well as the signature elastic/plastic phenomena in all PFC phenomenologies. These are important features for modelling solidification, CVD processes, and solidification shrinkage.

\section{Coupling the unified PFC model to thermal transport}
\label{UniHeat}

In this section, we return to the formalism from section~\ref{pfcheatcoupling}, and apply it to the unified model free energy derived in Section~\ref{newmodel}. The main motivation for this is to apply the thermo-density coupling to a quantitative  PFC type model that can be parameterized in $(P,\rho,T)$ space. We  re-write the free energy of \E{cdftmodel3}, making explicit the temperature dependence of the coefficients $a_m$,
\bea
\mathcal{F}&=&\int d\br \left[ \tilde{\mu}-\tilde{V}_{ext}- \left(\frac{\tau}{1-b}\right)\right] n_{mf} \nonumber\\
&+&\int d\br \,  \tau\left\{ \left(n_{mf}+1\right) \ln\left(\frac{\left(n_{mf}+1\right)(1-b)}{1-\left(n_{mf}+1\right)b}\right)-\frac{a}{\tau}n_{mf}^2 \right\} \nonumber\\
&-&\frac 1 2 B_x \! \int d\br\,  \tau(\br) e^{-\tau} n(\br) \mathcal{F}^{-1} \!\! \left\{ \int d\bk'  \hat{C}_{\bk^\prime} n(\bk') \right\} \nonumber\\
&+&\!\!\sum_{m=1}^{4} \frac{1}{m}  \left(\int d \br_1..d\br_ma_m(\tau)\zeta^{(m)}(\br_1,..,\br_m)n(\br_1)..n(\br_m) \! \right)\nonumber
\label{cdftmodel3_tau}
\eea

We next calculate the rescaled internal energy  $\tilde{e}=e/\bar{\rho}k_BT_0$, coarse grained on scales larger than the lattice constant, as was done in Section~\ref{cDFT_coupling} using the smoothing function  $\chi$. This gives, 
\bea
\tilde{e}&=&\chi*\left[f-\tau\frac{\delta \mathcal{F}}{\delta \tau}\right]=-a n_{mf}^2 \nonumber\\
&+&\!\! \frac 1 4 \! \sum_{m=2}^4 \chi\!\!*\!\! \left\{ \int d\br_2..d\br_m \left[   \right. \right. \label{engy_den} \\
&&\!\! \left.   \left( a_m-\tau a_m'\right) \zeta^m(\br_1,..,\br_m)n(\br_1)..(\br_m) \right] \bigg\}  \nonumber
\eea
Using the definition of the $\zeta^m$ functions and applying the delta functions within the $\zeta^m$ functions in \E{zeta_defII}, gives, after some manipulations and term collecting,
\be
\tilde{e}=-a n_{mf}^2 +\frac 1 4 \sum_{m=2}^4\left( a_m-\tau a_m'\right)\left\{ \chi*\left[n^m\right]- \left(\chi*\left[n\right]\right)^m \right\}
\ee
We next take the time derivative of $\tilde{e}$, yielding
\bea
\frac{\partial \tilde{e}}{\partial t}&=&\frac 1 4 \left\{ \sum_{m=2}^4\left(-\tau a_m'' \right)\left( \chi*\left[n^m\right]- \left(\chi*\left[n\right]\right)^m \right)\right\}\frac{\partial \tau}{\partial t}\nonumber\\
&+& \frac 1 4 \sum_{m=2}^4\left( a_m-\tau a_m'\right)\frac{\partial }{\partial t}\left\{ \chi*\left[n^m\right]- \left(\chi*\left[n\right]\right)^m \right\}\nonumber \\
&-&a \frac{\partial }{\partial t}\left(\chi*\left[n\right]\right)^2 \label{dedt}
\eea
We next proceed as in \E{rhs} and substitute $\tilde{e}$ into the dimensionless energy transport equation, 
\be
\frac{\partial \tilde{e}}{\partial t} =\mathcal{C}^*\nabla^2\tau,
\label{heatII}
\ee
where $\mathcal{C}^*= K/\bar{\rho}R^{2}k_B$. Namely, substituting \ref{dedt} for the left hand side of \E{heatII} yields
\bea
 && \frac{1}{4} \left\{ \sum_{m=2}^4\left(-\tau a_m'' \right)\left( \chi*\left[n^m\right] - \left(\chi*\left[n\right]\right)^m \right)\right\}\frac{\partial \tau}{\partial t} \nonumber\\
&&=\mathcal{C}^*\nabla^2\tau+a \frac{\partial }{\partial t}\left(\chi*\left[n\right]\right)^2 \label{finalfinalmodel} \\
&&- \frac 1 4 \sum_{m=2}^4\left( a_m-\tau a_m'\right)\frac{\partial }{\partial t}\left\{ \chi*\left[n^m\right]- \left(\chi*\left[n\right]\right)^m\right\} \nonumber
\eea
\EE{finalfinalmodel} is the analog of \E{PFC_heat_eq} for the more general  tmodel of \E{cdftmodel3}. Here, the source term and susceptibility factor pick up contributions from the moments of the PFC density field. The behaviour of this model and its coupling to temperature will be studied in future publications.

\bibliography{references.bib}
\bibliographystyle{unsrt}

\end{document}